%% file: 0_LLDBS.tex
\documentclass{article}
\pdfoutput=1
\usepackage{multirow} 
\usepackage[lined,linesnumbered,boxed]{algorithm2e}
\usepackage{color}
\usepackage{cite}
\usepackage{comment}
\usepackage{color}
\usepackage{tikz}
\usepackage{tkz-berge}
\usepackage{graphicx}
\usepackage{url}
\usepackage{subfig}
\usepackage{amsmath}

\usepackage[
        colorlinks=false,
        pdftitle={Boosting Multi-Core Reachability Performance with Shared Hash Tables},
        pdfauthor={Alfons Laarman, Jaco van de Pol, Michael Weber},
        pdfkeywords={model checking, multi-core, parallel, lockless, hash tables},
        pagebackref=true,
        naturalnames]{hyperref}
\title{Boosting Multi-Core Reachability Performance with Shared Hash Tables}

\author{{Alfons Laarman, Jaco van de Pol, Michael Weber}\\
	{\small\tt \{a.w.laarman,vdpol,michaelw\}@cs.utwente.nl}\\
	{\small Formal Methods and Tools, University of Twente, The Netherlands}
}

\begin{document}

\maketitle

\begin{abstract}
This paper focuses on data structures for multi-core reachability, which is a key component in model 
checking algorithms and other verification methods. A cornerstone of an efficient solution is the 
storage of visited states.  In related work, static partitioning of the state space was combined with 
thread-local storage and resulted in reasonable speedups, but left open whether improvements are 
possible.  In this paper, we present a scaling solution for shared state storage which is based on a 
lockless hash table implementation. The solution is specifically designed for the cache architecture of 
modern CPUs. Because model checking algorithms impose loose requirements on the hash table 
operations, their design can be streamlined substantially compared to related work on lockless hash 
tables.  Still, an implementation of the hash table presented here has dozens of sensitive performance 
parameters (bucket size, cache line size, data layout, probing sequence, etc.).  We analyzed their 
impact and compared the resulting speedups with related tools.  Our implementation outperforms two 
state-of-the-art multi-core model checkers (SPIN and DiVinE) by a substantial margin, while placing 
fewer constraints on the load balancing and search algorithms.
\end{abstract}

\newcounter{foot}
\setcounter{foot}{1}

\input{1_introduction}

\input{2_background}

\input{3_lockless_hashtable}
\input{4_experiments}

\input{5_conclusion}

\section{Acknowledgements}

We thank the chair of Computational Materials Science at UTwente for making their cluster available 
for our experiments. In particular, we thank Anton Starikov for aiding our experiments and at the same 
time configuring the cluster to facilitate them. We thank Petr Ro\v{c}kai and Ji\v{r}{\'\i} Barnat for the 
help and support they provided on the DiVinE toolkits.  We also thank the Linux Kernel developers for 
their help in tracing down the cause of the performance regression on newer kernels. This resulted in 
10\% improvements of the results.

\bibliography{9_LLDBS}
\bibliographystyle{plain}

\appendix
 \section{APPENDIX A -- SPEEDUPS}
This appendix contains detailed figures about per-model speedups with the different model checkers.
\begin{figure}[h]
\begin{center}
\includegraphics[width=0.9\textwidth]{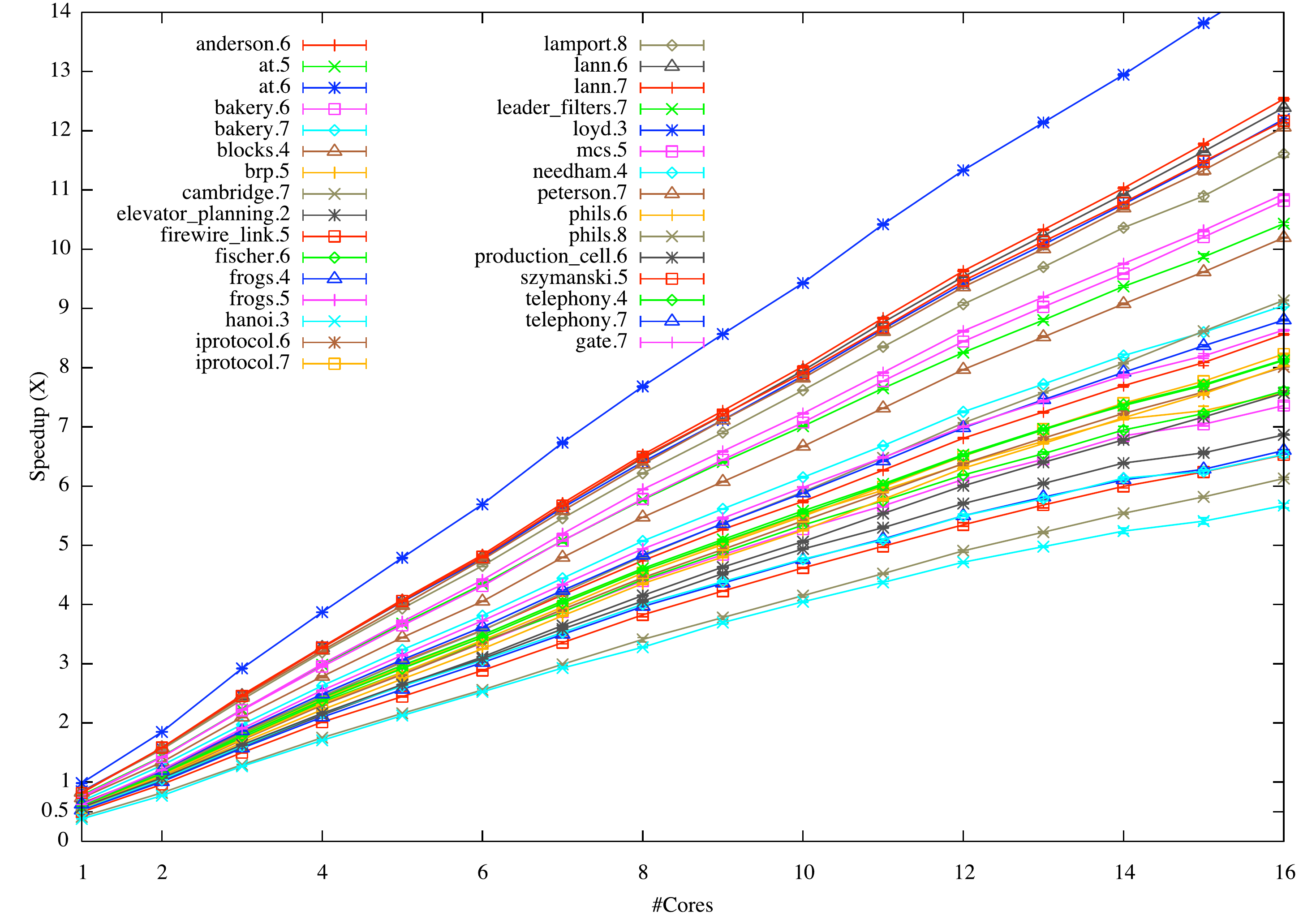}
\caption{Speedup of BEEM models with LTSmin}
\label{dve}
\end{center}
\end{figure}

\begin{figure}[h]
\begin{center}
\includegraphics[width=0.9\textwidth]{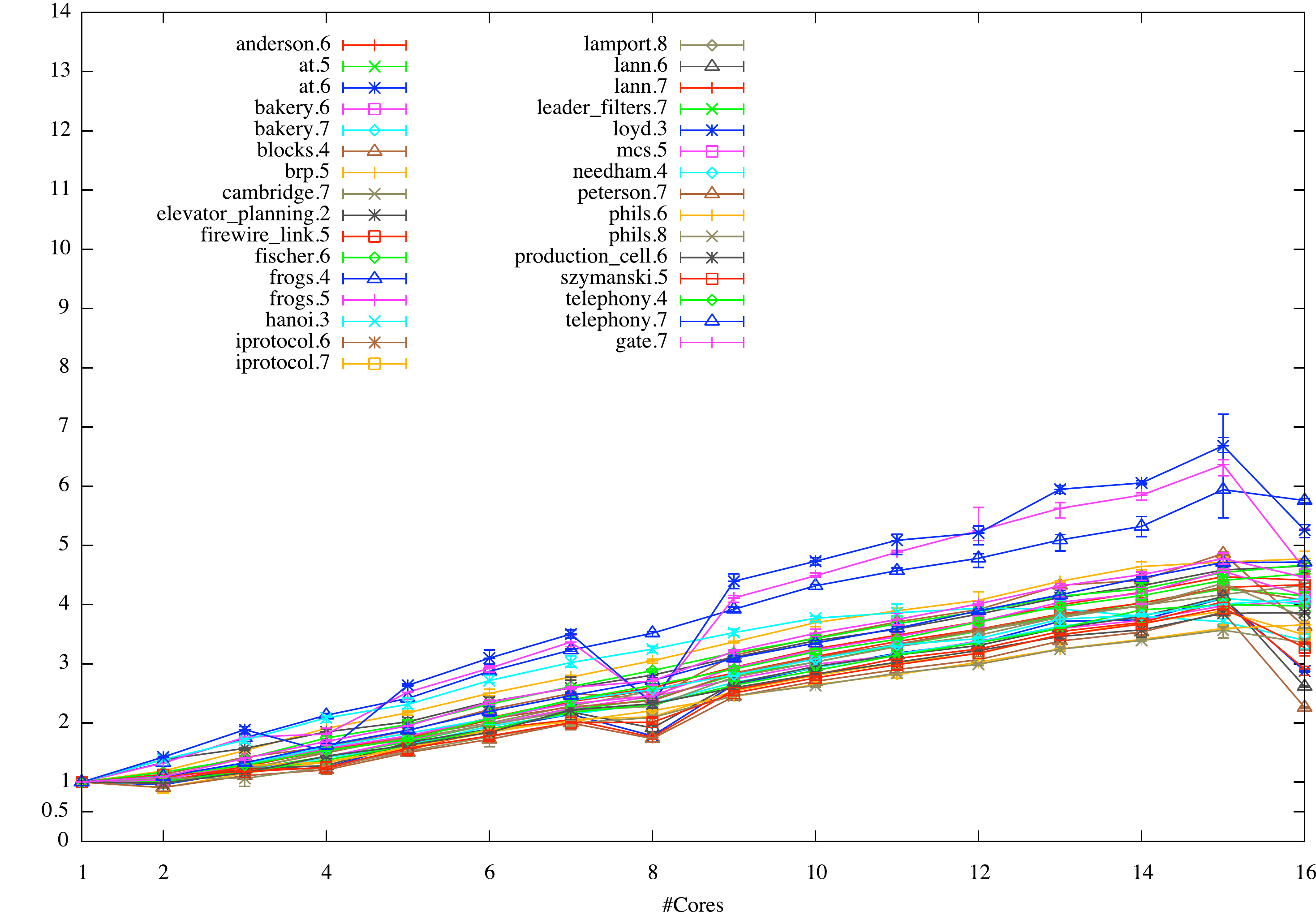}
\caption{Speedup of BEEM models with DiVinE 2.2}
\label{div2}
\end{center}
\end{figure}

\begin{figure}[h]
\begin{center}
\includegraphics[width=0.9\textwidth]{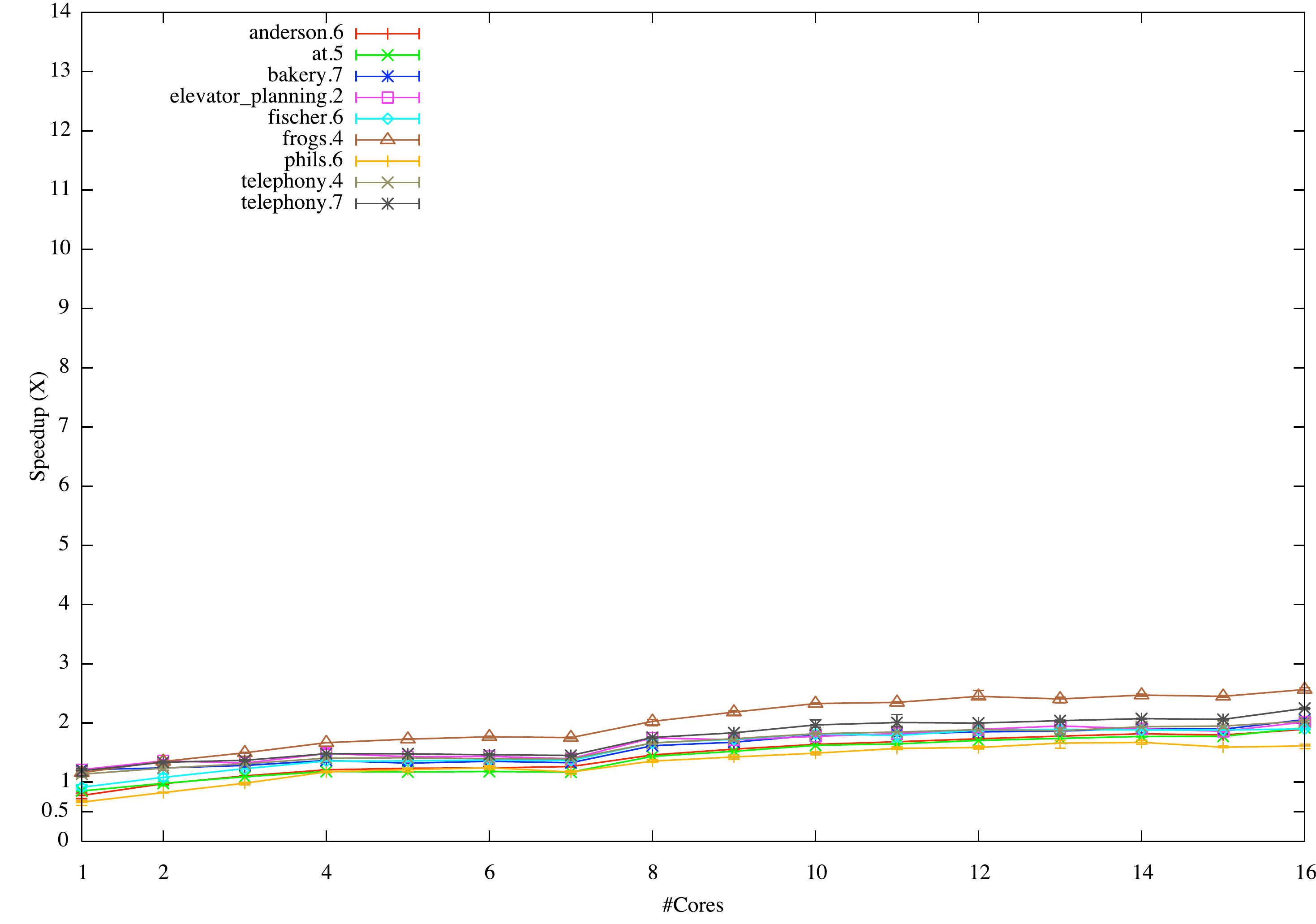}
\caption{Speedup of BEEM models with SPIN}
\label{div2-spin}
\end{center}
\end{figure}

\end{document}

%% file: 1_introduction.tex
\section{Introduction}

 Many verification problems are highly computational intensive tasks, which can benefit from extra 
speedups. Considering the recent trends in hardware, these speedups can only be delivered by 
exploiting the parallelism of the new multi-core processors.
 
 Reachability, or full exploration of the state space, is a subtask of many verification problems 
\cite{terminator,Brim:PDMC:2006}. In model checking, reachability has been parallelized in the past 
using distributed systems \cite{Brim:PDMC:2006}. 
With shared-memory systems, these algorithms can benefit from the low communication costs as has 
been demonstrated already \cite{BBR07}. In this paper, we show how state-of-the-art multi-core model 
checkers, like SPIN \cite{HB07} and DiVinE \cite{BBR07}, can be improved by a large factor (factor 
two  compared to DiVinE and four compared to SPIN) using a carefully designed concurrent hash table 
as shared state storage.

\textbf{Motivation.}  Holzmann and Bo\v{s}nacki used a shared hash table with fine grained locking in 
combination with the stack slicing algorithm in their multi-core extension of the SPIN model checker 
\cite{HB07, Holzmann20083}. This shared storage enabled the parallelization of many of the model 
checking algorithms in SPIN: safety properties, partial order reduction and reachability. Barnat et al. 
implemented the same method in the DiVinE model checker \cite{BBR07}. They also implemented the 
classic method of static state space partitioning, as used in distributed model checking 
\cite{Barnat200879}. They found the static partitioning method to scale better on the basis of 
experiments. The authors also mention that they were not able to investigate a potentially better 
solution for shared state storage, namely the use of a lockless hash table. 

\begin{figure}[tbp]
\begin{center}
 \subfloat[][Static partitioning]{
  \includegraphics[width=.5\textwidth]{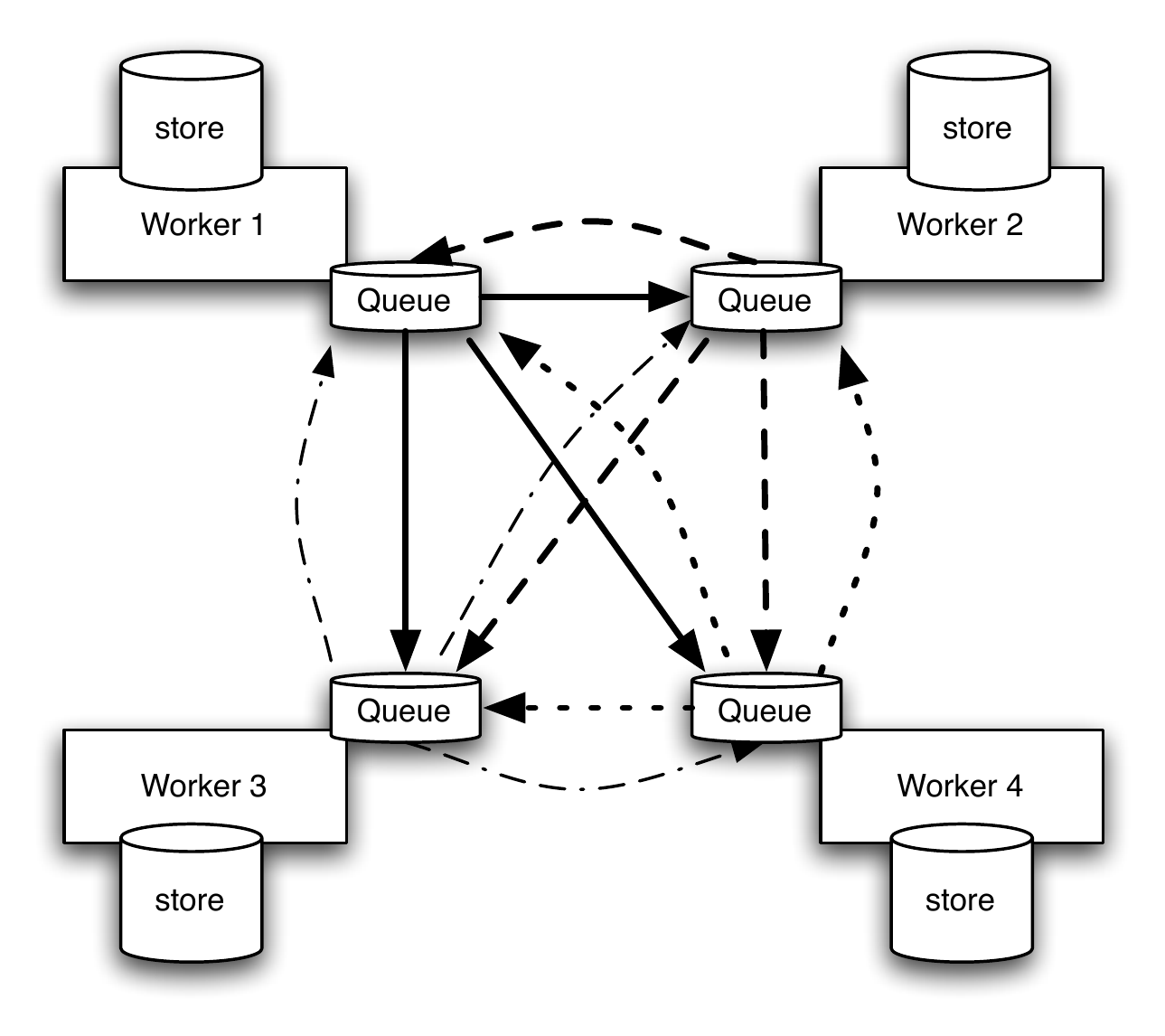}\label{fig:arch-static} }
 \subfloat[][Stack slicing]{
  \includegraphics[width=.5\textwidth]{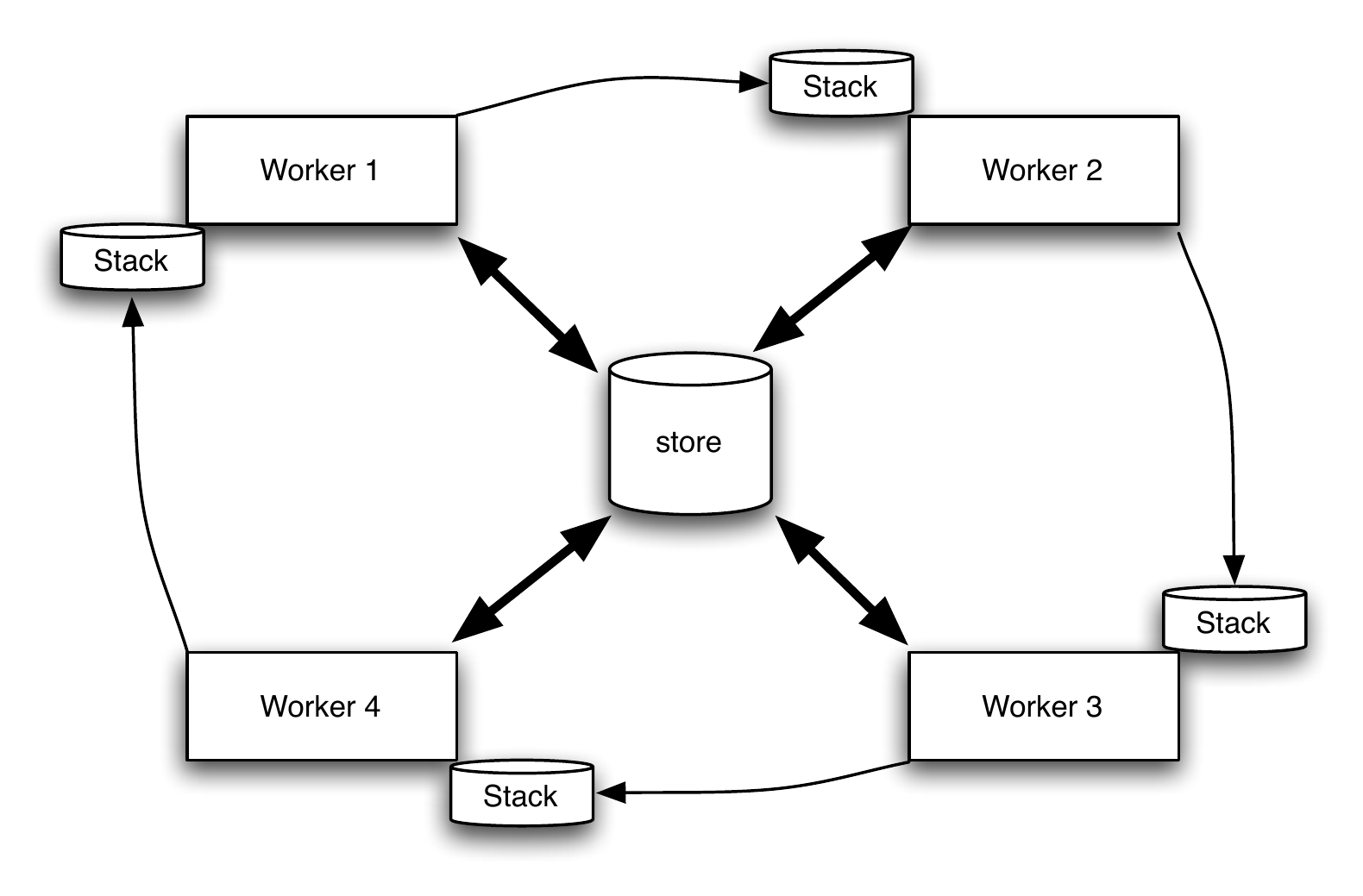} \label{fig:arch-slicing}}
  
 \subfloat[][Shared storage]{
  \includegraphics[width=.5\textwidth]{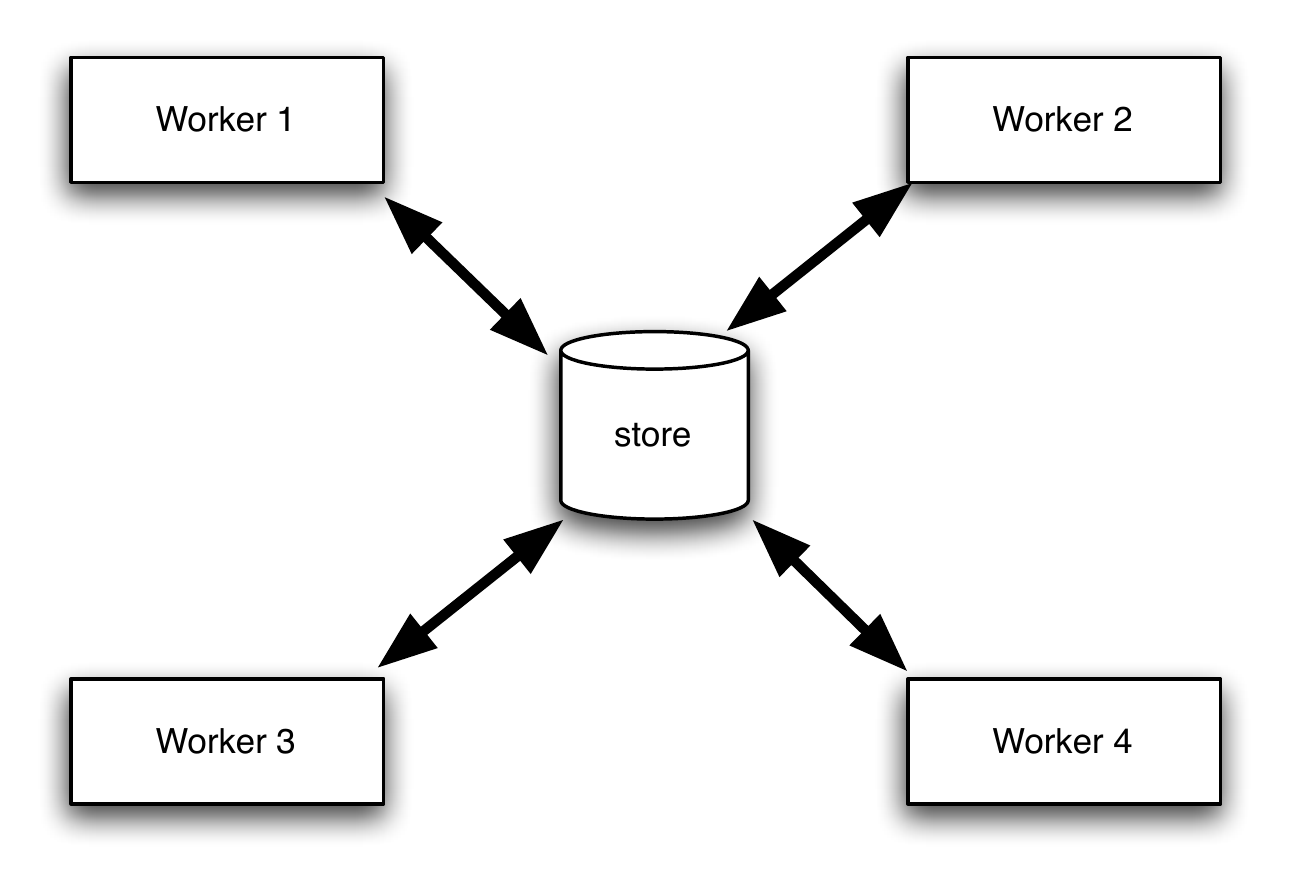} \label{fig:arch-shared}}
\end{center}
\caption{Different architectures for explicit model checkers}
\label{fig:archs}
\end{figure}

Fig.~\ref{fig:archs} shows the different architectures discussed thus far. The differences of these 
architectures is summarized in Table~\ref{tab:archs} and have been extensively discussed in 
\cite{Barnat200879}. From this, we deduce that the shared storage solution is both the simplest and the 
most flexible, in the sense that it allows any preferred exploration algorithm (except for true DFS which 
is inherently hard to parallelize). This includes (pseudo) DFS, which enables fast searches for 
deadlocks and error states \cite{RKV10}. SPIN already demonstrated this \cite{Holzmann20083}, but 
unfortunately does not show the desirable scalability (as we will demonstrate). Load balancing in SPIN 
is handled by the stack slicing algorithm \cite{Holzmann20083}, a specific case of explicit load 
balancing requiring DFS. In fact, any well-investigated load balancing solution \cite{sandersthesis} can 
be used and tuned to the specific environment, for example, to support heterogenous systems or BFS 
exploration. 
In conclusion, it remains unknown whether reachability, based on shared state storage, can scale.

\begin{table}[tp]
\caption{Differences between architectures}
\begin{center}
\begin{tabular}{|l|l|p{7cm}|}
\hline
\bf Arch.	& \bf Sync. point(s)	&\bf Pro's and Cons	\\ \hline
Fig \ref{fig:arch-static}	& Queue	& Local (\textit{cache efficient}) storage, \textit{static} load 
balancing, \textit{high} comm. costs, \textit{limited} to BFS	\\ \hline
Fig \ref{fig:arch-slicing}	& Shared store, stack	& Shared storage, \textit{static} load balancing, 
lower comm. costs, limited to (pseudo) DFS 	\\ \hline
Fig \ref{fig:arch-shared}	& Shared store(, queue)	& Shared storage, \textit{flexible} load balancing, 
no or fewer limits on exploration algorithm \\ \hline
\end{tabular}
\end{center}
\label{tab:archs}
\end{table}

Lockless hash tables and other efficient concurrent data structures are well known. Hash tables 
themselves have been around for a while now \cite{litwin80}. Meanwhile, variants like hopscotch 
\cite{HNM09} and cuckoo hashing \cite{PFR01} have been invented which optimize towards faster 
lookup times by reorganizing and/or indexing items upon insert. Hopscotch being particularly effective 
for parallelization, because of its localized memory behavior. Recently, also many lockless algorithms 
were proposed in this area \cite{opaddrht,CC07}. All of them providing a full set of operations on the 
hash table, including: lookup, insertion, deletion and resizing. They favor rigor and completeness over 
simplicity. None of these is directly suited to function as shared state storage inside a model checker, 
where it would have to deal with large state spaces, that grow linearly with a high throughput of new 
states. 

\textbf{Contribution.} The contribution of this paper is an efficient concurrent storage for states. 
This enables scaling parallel implementations of reachability  for any desirable exploration algorithm. 
To this end, the precise needs that parallel model checking algorithms impose on shared state storage 
are evaluated and a fitting solution is proposed given the requirements we identified. Experiments show 
that the storage scales significantly better than algorithm that uses static partitioning of the state 
space \cite{Barnat200879}, but also beats state-of-the-art model checkers. These experiments also 
contribute to the a better understanding of the performance of the latest versions of these model 
checkers (SPIN and DiVinE), which enables fair comparison.

With an analysis we also show that our design will scale beyond current state-of-the-art multi-core 
CPUs.

Furthermore, we contribute to the understanding of the more practical aspects of parallelization. With 
the proposed hash table design, we show how a small memory working set and taking into account the 
steep memory hierarchy can benefit the scalability on multi-core CPUs.

\textbf{Overview.} Section 2 gives some background on hashing, and parallel algorithms and 
architectures. Section 3 presents the lockless hash table that we designed for parallel state storage. 
But only after we evaluated the requirements that fast parallel algorithms impose on such a shared 
storage. This helps the understanding of the reasoning behind the design choices we made. The 
structure is implemented in a model checker together with parallel reachability algorithms. In Section 4 
then, the performance is evaluated against that of DiVinE 2 \cite{BBR09} and SPIN, which respectively 
use static partitioning and shared storage with stack slicing. A fair comparison can be made between 
the three model checkers on the basis of a set of models from the BEEM database that report the 
same number of states for both SPIN and DiVinE. Our model checker reuses the DiVinE next-state 
function and also reports the same number of states. We end this paper with some incentives we have 
for future work on the same topic and the conclusions (Section 5).


%% file: 2_background.tex
\section{Preliminaries}
\label{sec:preliminaries}

\textbf{Reachability in Model Checking} In model checking, a computational model of the system 
under verification (hardware or software) is constructed, which can then be used to compute all 
possible states of the system. The exploration of all states can be done symbolically using binary 
decision diagrams to represent sets of states and transitions or explicitly storing full states. While 
symbolic evaluation is attractive for a certain set of models, others, especially highly optimized ones, 
are better handled explicitly. We focus on explicit model checking in this paper.

Explicit reachability analysis can be used to check for deadlocks and invariants and also to store the 
whole state space and verify multiple properties of the system at once. A reachability run is basically 
a search algorithm which calls for each state the \textit{next-state} function of the model to obtain its 
successors until no new states are found. Alg.~\ref{alg:reach} shows this. It uses, e.g., a stack or a 
queue~$T$, depending on the preferred exploration order: depth or breadth first. The initial state $s_0$ 
is obtained from the model and put on~$T$. In the while loop on Line~1, a state is taken from~$T$, its 
successors are computed using the model (Line~3) and each new successor state is put into~$T$ 
again for later exploration. To determine which state is new, a set~$V$ is used usually implemented 
with a hash-table.

\begin{algorithm}[tp] 
\SetKwData{State}{state}\SetKwData{Succ}{succ}\SetKwData{Count}{count}\SetKwData{T}{T}
\SetKwFunction{NextState}{next-state}\SetKwFunction{Get}{get}\SetKwFunction{Put}{put}
\SetKwFunction{FindOrPut}{find-or-put}
\SetKwInOut{Input}{input}\SetKwInOut{Output}{output} 

\KwData{Sequence T = \{$s_0$\}, Set V = $\emptyset$} 
\BlankLine 
\While{\State $\gets$ $\T.\Get()$}{ 
	\Count $\gets$ 0\;
	\For{\Succ \bf{in} \NextState(\State)}{
       	\Count $\gets$ \Count + 1\;
	\If(){V.\FindOrPut(\Succ)} { 
         $\T.\Put(\Succ)$\;
       }    }
   \If(){$0 == \Count$} { 
	//DEADLOCK, print trace..
   }
} 

\caption{Reachability analysis} 
\label{alg:reach} 
\end{algorithm} 

Possible ways to parallelize Alg.~\ref{alg:reach} have been discussed in the introduction. A common 
denominator of all these approaches is that the strict BFS or DFS order of the sequential algorithm is 
sacrificed in favor of local stacks or queues (fewer contention points).
When using a shared state storage (in a general setup or with stack slicing), a thread-safe set $V$ is 
required, which will be discussed in the following section.

\paragraph*{Load Balancing.} A naive parallelization of reachability is a limited sequential BFS 
exploration and then handing of the found states to several threads that start  executing Alg.~
\ref{alg:reach} ($T =$ \{\textit{part of BFS exploration}\} and $V$ is shared). This is called \textit{static 
load balancing}. For many models this will work due to common diamond-shaped state spaces. 
Models with synchronization points, however, have flat helix-shaped state spaces, so threads will run 
out of work when the state space converges. A well-known problem that behaves like this is the 
Tower of Hanoi puzzle; when the smallest disk is on top of the tower only one move is possible at that 
time.

Sanders \cite{sandersthesis} describes \textit{dynamic load balancing} in terms of a problem $P$, a 
(reduction) operation $work$ and a $split$ operation. $P_{root}$ is the initial problem (in the case of 
Alg~\ref{alg:reach}: T=\{initial\_state\}). Sequential execution takes $T_{seq} = T(P_{root})$ time units. 
A problem is (partly) solved when calling $work$(P, t), which takes min(t, T(P)) units of time. For our 
problem, work(P,t) is the reachability algorithm with $P=T$ and $t$ has to be added as an extra input, 
that limits the number of iterations of the while loop (line~1). When threads become idle, they can poll 
others for work. On which occasion, the receiver will split its own problem ($split(P)=\{P_1,P_2\} 
\rightarrow T(P) = T(P_1) + T(P_2)$) and send one of the results to the polling thread. We 
implemented synchronous random polling and did not notice real performance overhead compared to 
no load balancing.

\paragraph*{Parallel architectures}
We consider multi-core and multi-CPU systems. A common desktop PC has a multi-core CPU, 
allowing for quick workspace verification runs with a model checker. The typical server, nowadays, 
has multiple of such CPUs on one board. Ideal for performance model checking, but more complex to 
program due to the more complex bus structure with different latencies and non-uniform memory 
access.

The cache coherence protocol ensures that each core has a global view of the memory. While the 
synchronization between caches may cost little, the protocol itself causes overhead. When all cores 
of a many-core system are under full load and communicate with each other, the data bus can be 
easily exhausted.
The cache coherence protocol cannot be preempted. To efficiently program these machines, few 
options are left. One way is to completely partition the input as done in \cite{Barnat200879}, this 
ensures per-core memory locality at the cost of increased inter-die communication. An improvement 
of this approach is to pipeline the communication using ring buffers, this allows prefetching (explicit or 
hardwired). This is done in \cite{MP09}. If these options are not possible for the given problem, the 
option that is left is to minimize the \textit{memory working set} of the algorithm \cite{opaddrht}. We 
define the memory working set as the number of different memory locations that the algorithm 
updates in the time window that these usually stay in local cache. A small working set minimizes 
coherence overhead.
 
\paragraph*{Locks} are used for mutual exclusion and prevent concurrent accesses to a critical 
section of the code. For resources with high contention, locks become infeasible. Lock proliferation 
improves on this by creating more locks on smaller resources. Region locking is an example of this, 
where a data structure is split into separately locked regions based on memory locations. However, 
this method is still infeasible for computational tasks with high throughput. This is caused by the fact 
that the lock itself introduces another synchronization point; and synchronization between processor 
cores takes time. 

\paragraph*{Lock-free algorithms} (without mutual exclusion) are preferred for high-throughput 
systems. Lockless algorithms postpone the commit of their result until the very end of the 
computation.
This ensures maximum parallelism for the computation, which may however be wasted if the commit 
fails. That is, if meanwhile another process committed something at the same memory address. In 
this case, the algorithm need so ensure progress in a different way. This can be done in varyingly 
complicated ways and introduces different kinds of lockless implementations: an algorithm is 
considered lock-less if there is guaranteed system-wide progress; i.e. always one thread can 
continue. A wait-free algorithm also guarantees per-thread progress. 

Many modern CPUs implement an ``Compare\&Swap'' operation (CAS) that ensures atomic memory 
modification, while at the same time preserving data consistency if used in the correct manner. For 
data structures, it is easy to construct lockless modification algorithms by reading the old value in the 
structure, performing the desired computation on it and writing the result back using CAS. If the latter 
returns true, the modification succeeded, if not, the computation needs to be redone with the new 
value or some other collision resolution should be applied.

\begin{algorithm}[tp]
\begin{tabbing}
\textit{Pre}: \=$word \neq \textbf{null}$\\
\textit{Post}: \>$(*word_{pre} = testval \Rightarrow *word_{post} = testval) \land $\\
\>$ (*word_{pre} \neq testval \Rightarrow *word_{post} = oldval) $\\
\textbf{atomic} bool compare\_and\_swap (int *word, int testval, int newval)
\end{tabbing}
\caption{``Compare\&Swap'' specification with C pointer syntax} 
\label{alg:hash} 

\end{algorithm} 

If a CAS operation fails (returns false), the computational efficiency decreases. However, in the 
normal case, where the time between reading a value from the data structure and writing it back is 
typically small due to short computations, collisions will rarely occur. So lockless algorithms can 
achieve a high level of concurrency there. Although it should not go unmentioned that an instruction 
like CAS easily costs 100--1000 of instruction cycles depending on the CPU architecture. Thus, 
abundant use defies the purpose of lockless algorithms.

\paragraph*{Quantifying parallelism} is usually done by normalizing performance gain with regard to 
the sequential run: 
$S = T_{seq} / T_{par}$ 
Linear speedups grow proportional to the number of cores and indicate that an algorithm scales.
The efficiency gives a measure of how much the extra computing power materialized into speedups: 
$E = (N \times T_{par}) / T_{seq}$, where N is the number of cores. 

$E$ can be larger than 1, in which case we speak of super-linear speedups. They can occur in several 
situations:
\begin{itemize}
\item more cache becomes available with the extra CPU reducing the amount of lookups in secondary 
memory,
\item a more efficient algorithm, than the sequential one, is found, or
\item the parallelization introduces randomization, which exponentially decreases the likelihood of 
taking the longest path in a search problem \cite{speedup}.
\end{itemize}
A parallel search for error states or deadlocks, could thus greatly benefit from a shared state storage, 
since it allows depth-preferred exploration \cite{RK88}.

\paragraph*{Hashing} is a well-studied method for storing and retrieving data with time complexity 
$O(1)$ \cite{litwin80}. A hash function \textit{h} is applied to the data, yielding an index in an array of 
buckets that contain the data or a pointer to the data. Since the domain of data values is usually 
unknown and much larger than the image of \textit{h}, hash collisions occur when $h(D_1) = h(D_2)$ 
with $D_1 \ne D_2$. Structurally, collisions can be resolved either by inserting lists in the buckets 
(chaining) or by probing subsequent buckets (open addressing). Algorithmically, there is a wealth of 
options to maintain the ``chains'' and calculate subsequent buckets \cite{Cormen}. The right choice to 
make depends entirely on the requirements dictated by the algorithms that use the hash table.

Generalized cuckoo hashing employs $d>1$ independent hash functions and buckets of size $n$ 
\cite{PFR01}. Elements are inserted in bucket $h_i$ ($i \in \{1,\dots,d\}$) which is the least full. It is 
well known that this method already improves the distribution exponentially compared to the case 
where $d=1$ \cite{balanced}. But cuckoo hashing also ensures constant time lookups by recursively 
reassigning all elements in the bucket to another $h_i$,  if a fixed file-ratio $m/n$ for a bucket is 
reached. The process is recursive, because the rearrangement may trigger other buckets to reach the 
threshold $m/n$. If the rearrangement does not terminate, a resize of the table is triggered.

Because cuckoo hashing requires $d$ independent memory references for each operation, it is hard to 
parallelize efficiently on general purpose hardware. Therefore, hopscotch hashing was introduced 
\cite{HNM09}. It keeps $d=1$, but uses the same reordering principle to move elements within a fixed 
size of their primary location. The fixed size is usually chosen within the range of a cache line, which 
is the minimum block size that can be mapped to the CPU's L1 cache.


%% file: 3_lockless_hashtable.tex
\section{A Lockless Hash Table}
In principle, Alg.~\ref{alg:reach} seems easy to parallelize, in practice it is difficult to do this efficiently 
because of the memory intensive behavior, which becomes more obvious when looking at the 
implementation of $V$. In this section, we present an overview of the options in hash table design. 
There is no silver bullet design and individual design options should be chosen carefully considering 
the requirements stipulated by the use of the hash table. Therefore, we evaluate the demands that the 
parallel model checking algorithms place on the state storage solution. We also mention additional 
requirements that come from used hardware and software systems.
Finally, a specific hash table design is presented.

\subsection{Requirements on the State Storage}
Our goal is to realize an efficient shared state storage for parallel model checking algorithms. 
Traditional hash tables associate a piece of data to a unique key in the table. In model checking, we 
only need to store and retrieve states, therefore the key is the data of the state. Henceforth, we will 
simply refer to it as \textit{data}. Our specific model checker implementation introduces additional 
requirements, discussed later. First, we list the definite requirements on the state storage: 

\begin{itemize}
\item The storage needs only one operation: \textit{find-or-put}. This operation inserts the state vector 
if its not found or yields a positive answer without side effects. This operation needs to be 
concurrently executable to allow sharing the storage among the different processes. Other operations 
are not necessary for reachability algorithms and their absence simplifies the algorithms thus lowering 
the strain on memory and avoiding cache line sharing. This in sharp contrast to standard literature on 
concurrent hash tables, where often a complete solution is presented optimizing them for more 
generalized access patterns \cite{opaddrht,CC07}.
\item The storage should not require memory allocation during operation, for the obvious reasons that 
this behavior would increase the \textit{memory footprint}.
\item The use of pointers on a per-state basis should be avoided. Pointers take a considerable 
amount of memory when large state spaces are explored (more than $10^8$ states are easily 
reachable with todays model checking systems), especially on 64-bit machines. In addition, pointers 
increase the \textit{memory footprint}.
\item The time efficiency of \textit{find-or-put} should scale with the number of processes executing it. 
Ideally, the the individual operations should -- on average -- not be slowed down by other operations 
executing at the same time ensuring close-to linear \textit{speedup}. Many hash table algorithms have 
a large \textit{memory footprint} due to their probe behavior or reordering behavior upon inserts. They 
cannot be used concurrently under high-throughputs as is the case here.

\end{itemize}

Specifically, we do not require the storage to be resizable. The available memory on a system can 
safely be claimed by the table, since the most of it will be used for it anyway. This requirement is 
justifiable, because exploring larger models is more lucrative with other options: (1) disk-based model 
checkers \cite{JE10} or (2) bigger systems.
In sequential operation and especially in presence of a delete operation (shrinking tables), one would 
consider resizing for the obvious reason that it improves locality and thus cache hits, in a concurrent 
setting, however, these cache hits have the opposite effect of causing the earlier described cache line 
sharing among CPUs or: dirty caches. We tested some lockless and concurrent resizing mechanisms 
and observed large decreases in performance.

Furthermore, the design of the used model checker also introduces some specific requirements:

\begin{itemize}
\item The storage stores integer arrays or vectors of known length vector-size. This is the encoding 
format for states employed in the model checker.
\item The storage should run on common x86 architectures using only the available (atomic) 
instructions.
\end{itemize}

While the compatibility with the x86 architecture allows for concrete analysis, the applicability is not 
limited to it. Lessons learned here are transferrable to other similar settings where the same memory 
hierarchy is present and the atomic operations are available.

\subsection{Hash Table Design}
The first thing to consider is the type of hash table to choose. Cuckoo hashing is an unlikely 
candidate, since it requires updates on many locations upon inserts. Using such an algorithm would 
easily result in a high level of cache line sharing and an exhaustion of the processor bus with cache 
coherence traffic as we witnessed in early attempts at creating a state storage.

Hopscotch hashing could be considered because it combines a low \textit{memory footprint} with 
constant lookup times even under higher load factors. But among the stated requirements, scalable 
performance is the most crucial. Therefore, we choose a simpler design which keeps the memory 
footprint as low as possible. Later we will analyze the difference with hopscotch hashing.
These considerations led us to the following design choices:

\begin{itemize} 
\item \textbf{Open addressing} is preferred, since chaining would incur in-operation memory allocation 
or pre-allocation at different addresses leading to more cache line sharing.
\item \textbf{Walking-the-line} is a name we gave to linear probing on the cache line followed by 
double hashing as also employed in \cite{CC07, HNM09}. The first allows a core to benefit mostly 
from one prefetched cache line, while the second mode realizes better distribution.
\item \textbf{Separated data} (vectors) in an indexed array $(buckets * |vector|)$ ensures that the 
bucket array stays short and subsequent probes can be cached.
\item \textbf{Hash memoization} speeds up probing, by storing the hash (or part of it) in the bucket. 
This prevents expensive lookups in the data array \cite{CC07}.
\item \textbf{A $2^n$ sized table} gives good probing distribution and can avoid the expensive modulo 
instruction, because the $n$ least-significant bits of the hash can be used as an index in the bucket 
array. 
\item \textbf{Lockless} operation using a dedicated value to indicate free places in the hash array 
(zero for example). One bit of the hash can be used to indicate whether the vector was already written 
to the data array or whether this is still in progress \cite{CC07}.
\item \textbf{Compare-and-swap} is used as the atomic function on the buckets, which are now in 
either of the following distinguishable states: \textit{empty}, \textit{being written} and \textit{complete}.
\end{itemize}    

The required table size may become a burden when aiming to utilize the complete memory of the 
system. To allow different sized tables, constant sized division can be used \cite{delight}.

\subsection{Algorithm}\label{sec:algo}

Alg.~\ref{alg:findorput} shows the \textit{find-or-put} operation. Buckets are represented by the 
$Bucket$ array, the separate data by the $Data$ array and hash functions used for double hashing by 
$h_i$. Probing continues indefinitely (Line~4) or until either a free bucket is found for insert 
(Line~8--10) or the data is found to be in the hash table (Line~15--17). The for loop on Line 5 handles 
the walking-the-line sequential probing behavior (Alg.~\ref{alg:walk}). The other code inside this for 
loop handles the synchronization among threads. It requires explanation.

\begin{figure}[height=1cm]
\centering
\begin{tikzpicture}[scale=.7, transform shape]
    \tikzstyle{VertexStyle}=[shape        = circle,
	                              ball color      = white!500!gray,
					minimum size = 45pt,
					draw]
\Vertex[x=0, y=1.7]{empty}
\Vertex[x=4, y=1.7]{write $x_1$}
\Vertex[x=8, y=1.7]{done $x_1$}
\Vertex[x=4, y=0]{write $x_2$}
\Vertex[x=8, y=0]{done $x_2$}
\Vertex[x=4, y=-1.7,style=dashed]{write}
\Vertex[x=8, y=-1.7,style=dashed]{done}
\Vertex[x=4, y=-3.4]{write $x_n$}
\Vertex[x=8, y=-3.4]{done $x_n$}
\Edge[style=->](empty)(write $x_1$)
\Edge[style=->](write $x_1$)(done $x_1$)
\Edge[style=->](empty)(write $x_2$)
\Edge[style=->](write $x_2$)(done $x_2$)
\Edge[style={dashed,->}](empty)(write)
\Edge[style={dashed,->}](write)(done)
\Edge[style=->](empty)(write $x_n$)
\Edge[style=->](write $x_n$)(done $x_n$)
 \end{tikzpicture}

\caption{State diagram of buckets} 
\label{fig:bucket_states} 
\end{figure}

Buckets store memoized hashes and the write status bit of the data in the $Data$ array. Fig.~
\ref{fig:bucket_states} shows the possible states of the buckets in the hash table, memoized hashes 
are depicted as $x_i$, and the write state of the data array is either  $write$ (Line~7), meaning that 
the data is being written or $done$ (Line~9), meaning that the write is completed. Whenever a write 
started for a hash value $x_1$ the state of the bucket can never become empty again, nor can it be 
used for any other hash value. This ensures that the probe sequence remains deterministic and 
cannot be interrupted.

\begin{algorithm}[tp] 
\SetKwData{Num}{num}\SetKwData{HashMem}{hash\_memo}\SetKwData{Index}{index}
\SetKwData{Candidate}{candidate}\SetKwData{Vector}{vector}\SetKwData{Bucket}{Bucket}
\SetKwData{Data}{Data}\SetKwData{Size}{size}
\SetKwData{Idx}{i}
\SetKwFunction{WalkTheLineFrom}{walkTheLineFrom}
\SetKwFunction{CAS}{CAS}
\SetKwFunction{Hash}{hash}
\SetKwInOut{Input}{input}\SetKwInOut{Output}{output} 

\KwData{size, Bucket[size], Data[size]} 
\Input{vector} 
\Output{seen} 
\BlankLine 

\Num$\gets$ 1\;
\HashMem$\gets {\Hash}_{\Num}(\Vector)$\;
\Index $\gets \Hash \mod \Size$\;
\newcommand\bstate[1]{\langle{#1}\rangle}

\While{true}{ 
\For{\Idx \bf{in} $\WalkTheLineFrom(\Index)$}{
	\If(){empty = $\Bucket[\Idx]$} { 
		\If(){$\CAS(\Bucket[\Idx], empty, \bstate{\HashMem,write})$} { 
			$\Data[\Idx] \gets \Vector$\;
			$\Bucket[\Idx] \gets$ $ \bstate{\Hash, done}$\;
			\textbf{return false}\;
		}
	}
	\If(){$\HashMem$  = $\Bucket[\Idx]$} { 
		\bf{while} {$\bstate{-, write}$  = $\Bucket[\Idx]$} \bf{do} \textit{..wait..} \bf{done} \\  
		\If(){$\bstate{-,done} $ $=\Bucket[\Idx] \wedge \Data[\Idx]=\Vector$ } { 
			return true\label{wait}\;
		}
	}
} 
\Num$\gets$ \Num + 1\;
\Index $\gets \Hash_{\Num}(\Vector) \mod \Size$\;
} 

\caption{The \textit{find-or-put} algorithm} 
\label{alg:findorput} 
\end{algorithm} 

\begin{algorithm}[tp] 
\SetKwData{X}{i}\SetKwData{Index}{index}\SetKwData{Walk}{Walk}
\SetKwData{CLS}{cache\_line\_size}\SetKwData{Start}{start}
\SetKwInOut{Input}{input}\SetKwInOut{Output}{output}

\KwData{\CLS, $\Walk[\CLS]$}
\Input{\Index} 
\Output{\Walk[]} 
\BlankLine 

$\Start \gets \lfloor \Index / \CLS \rfloor \times$ \CLS\;
\For {\X $\gets$ 0 \KwTo $\CLS-1$ } {
	$\Walk[\X] \gets (\Start + \Index +\X) \mod{\CLS}$\;
}

\caption{Walking the (cache) line} 
\label{alg:walk} 
\end{algorithm} 

Several aspects of the algorithm guarantee correct lock-less operation:
\begin{itemize}
\item The CAS operation on Line~7 ensures that only one thread can claim an empty bucket, marking 
it as non-empty with the hash value to memoize and with $write$.
\item The while loop on Line~14 waits until the write to the data array has been completed, but it only 
does this if the memoized hash is the same as the hash value for the vector (Line~13).
\end{itemize}

So the critical synchronization between threads occurs when both try to write to an empty slot. The 
CAS operation ensures that only one will succeed and the other can carry on with the probing, either 
finding another empty bucket or finding the state in another bucket.
This design can be seen as a lock on the lowest possible level of granularity (individual buckets), but 
without a true locking structure and associated additional costs. The algorithm implements the ``lock''  
as while loop, which resembles a spinlock (Line~14). Although it could be argued that this algorithm is 
therefore not lockless, it is possible to implement a resolution mechanism in the case that the 
``blocking'' thread dies or hangs ensuring local progress (making the algorithm wait-free). This is 
usually done by making each thread fulfill local invariants, whenever they could not be met by other 
threads \cite{amp}. It can be done by directly looking into the buffer of the writing thread, whenever the 
``lock'' is hit, and finish the write locally when the writing thread died. Measurements showed, 
however, that the ``locks'' are rarely hit under normal operation, because of the hash memoization.

The implementation of the described algorithm requires exact guarantees by the underlying memory 
model. Reordering of operations by compilers and processors needs to be avoided across the 
synchronization points or else the algorithm is not correct anymore. It is, for example, a common 
optimization to execute the body of an if statement before the actual branching instruction. This 
enables speculative execution, keeping the processor pipeline busy as long as possible. This would 
be a disastrous reordering when applied to Line~8 and Line~9: the actual write would happen before 
the bucket is marked as full, allowing other threads to write to the same bucket.  

The Java Memory Model makes precise guarantees about the possible commuting of memory reads 
and writes, by defining a partial ordering on all statements that effect the concurrent execution model~
\cite[Sec.~17.4]{JavaSpec}. A correct implementation in Java should declare the bucket array as a 
volatile variable and use the \texttt{java.util.concurrent.atomic} for atomic references and CAS. A 
C~implementation is more difficult, because the ANSI~C99 standard does not state any requirements 
on the memory model. The implementation would depend on the combination of CPU architecture and 
compiler. Our implementation uses gcc with x86 64-bit target platforms.  A built-in function of gcc is 
used for the CAS operation and reads and writes from and to buckets are marked volatile.

Alg.~\ref{alg:findorput} was modeled in PROMELA and checked with SPIN. One bug concerning the 
combination of write bit and memoized hash was found and corrected.


we intend to deliver a more thorough analysis at a later time.  Table~\ref{tab:complexities} shows the 
how expected number of probes depends for successful and unsuccessful lookups (reads and writes) 
is dependent on the fillrate.

%% file: 4_experiments.tex
\section{Experiments}

\subsection{Methodology}

We implemented the the hash table of the previous section in our our own model checking toolset 
LTSmin, which will be discussed more in later sections. For the experimental results it suffices to 
know that LTSmin reuses the next-state function of Divine. Therefore, a fair comparison with DiVinE 
can be made. We also did experiments with the latest multi-core capable version of the model 
checker Spin \cite{HB07}. For all the experiments, the reachability algorithm was used, because it 
gives the best evaluation of the performance of the state storage. Spin and DiVinE 2 use static state 
space partitioning \cite{HB07, Barnat200879}, which is a breadth-first exploration with a hash function 
that assigns each successor state to the queue of one thread. The threads then use their own state 
storage to check whether states are new or seen before. 

All model checkers were configured for maximum performance in reachability. For DiVinE and LTSmin 
this meant that we used compiled (DiVinE also contains a model interpreter), with an initial hash table 
size large enough to contain the models state space, while not triggering a resize of the table. For 
SPIN, we turned of all analysis options, state compression and other advanced options. To compile 
spin models we used the flags: \texttt{-O3} \texttt{-DNOCOMP} \texttt{-DNOFAIR} \texttt{-
DNOREDUCE} \texttt{-DNOBOUNDCHECK} \texttt{-DNOCOLLAPSE} \texttt{-DNCORE=}$N$  \texttt{-
DSAFETY} \texttt{-DMEMLIM=100000}; To run the models we used the options: \texttt{-m10000000} 
\texttt{-c0} \texttt{-n} \texttt{-w28}.
The machines we used to run experiments are AMD Opteron~8356 16-core servers running a patched 
Linux~2.6.32 kernel\footnote[\value{foot}]{During the experiments we found large performance 
regression in the newer 64bit kernels, which where solved with the help of the people from the Linux 
Kernel Mailing List: \\ \url{https://bugzilla.kernel.org/show_bug.cgi?id=15618}}. All programs were 
compiled using gcc~4.4 in 64-bit mode with maximum compiler optimizations~(\texttt{-O3}).  

\addtocounter{foot}{1}

A total of 31~models randomly chosen from the BEEM database \cite{beem} have been used in the 
experiments (only too small and too large models have been filtered out). Every run was repeated at 
least four times, to exclude any accidental mis-measurements.
Special care has been taken to keep all the parameters across the different model checkers the 
same. Especially SPIN provides a rich set op options with which models can tuned to perform 
optimal. Using these parameters on a per-model basis could give faster results than presented here, 
just like if we optimized Divine for each special case. It would, however, say little about the scalability 
of the core algorithms.

\begin{table}[]
\caption{Benchmark suite for DiVinE, LTSmin and SPIN (*)}
\begin{center}
\begin{tabular}{| l | l | l | l | l |}
\hline
anderson.6 *	&at.5 *	&at.6	&bakery.6				 \\ \hline
bakery.7 * & blocks.4	&brp.5	&cambridge.7		 \\ \hline 
frogs.5	&hanoi.3	&iprotocol.6 & elevator\_planning.2 	*	 \\ \hline
firewire\_link.5 & fischer.6 *	&frogs.4 *	& iprotocol.7		\\ \hline
lamport.8	&lann.6	&lann.7	&leader\_filters.7 \\ \hline
loyd.3	&mcs.5	&needham.4	&peterson.7		\\ \hline
phils.6 * & phils.8	&production\_cell.6	&szymanski.5 \\ \hline
telephony.4 * &telephony.7 *  &	train\_gate.7 & \\ \hline

\end{tabular}
\end{center}
\label{tab:models}
\end{table}

Therefore, we decided to leave all the parameters the same for all the models. Resizing of the state 
storage could be avoided in all cases by increasing its initial size. This means that small models use 
the same large state storage as large models.

\subsection{Results}

\begin{figure}[h!]
\begin{center} 
 \includegraphics[width=.7\textwidth]{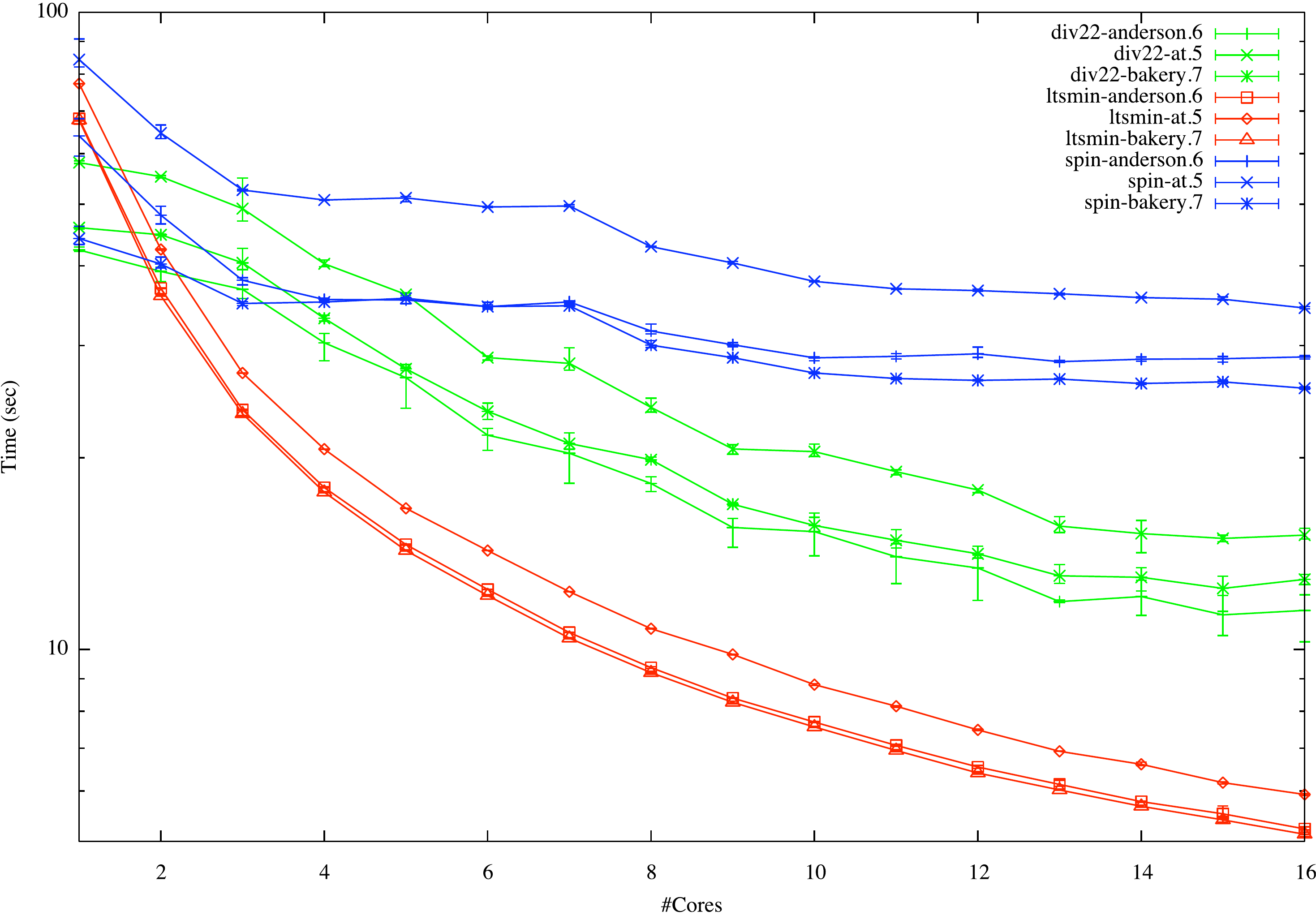}
\end{center}
 \caption{Runtimes of three BEEM models with SPIN, LTSmin and DiVinE 2}
\label{fig:runtimes}
\end{figure}

Representing the results of so many benchmark runs in a concise and clear manner can hardly be 
done in a few graphs. Figure~\ref{fig:runtimes} shows the run times of only three models for all model 
checkers. We see that DiVinE is the fastest model checker for sequential reachability. Since the last 
published comparison between DiVinE and SPIN \cite{BBR07}, DiVinE has been improved with a 
model compiler and a faster next-state function\footnote[\value{foot}]{See release notes version 2.2}. 
The figure shows that these gains degraded the scalability, which is a normal effect as extensively 
discussed in \cite{HB07}. SPIN is only slightly slower than DiVinE and shows the same linear curve 
but with a more gentle slope. We suspect that the gradual performance gains are caused by the cost 
of the inter-thread communication. 
\addtocounter{foot}{1}

\begin{figure}[h!]
\begin{center} 
 \includegraphics[width=\textwidth]{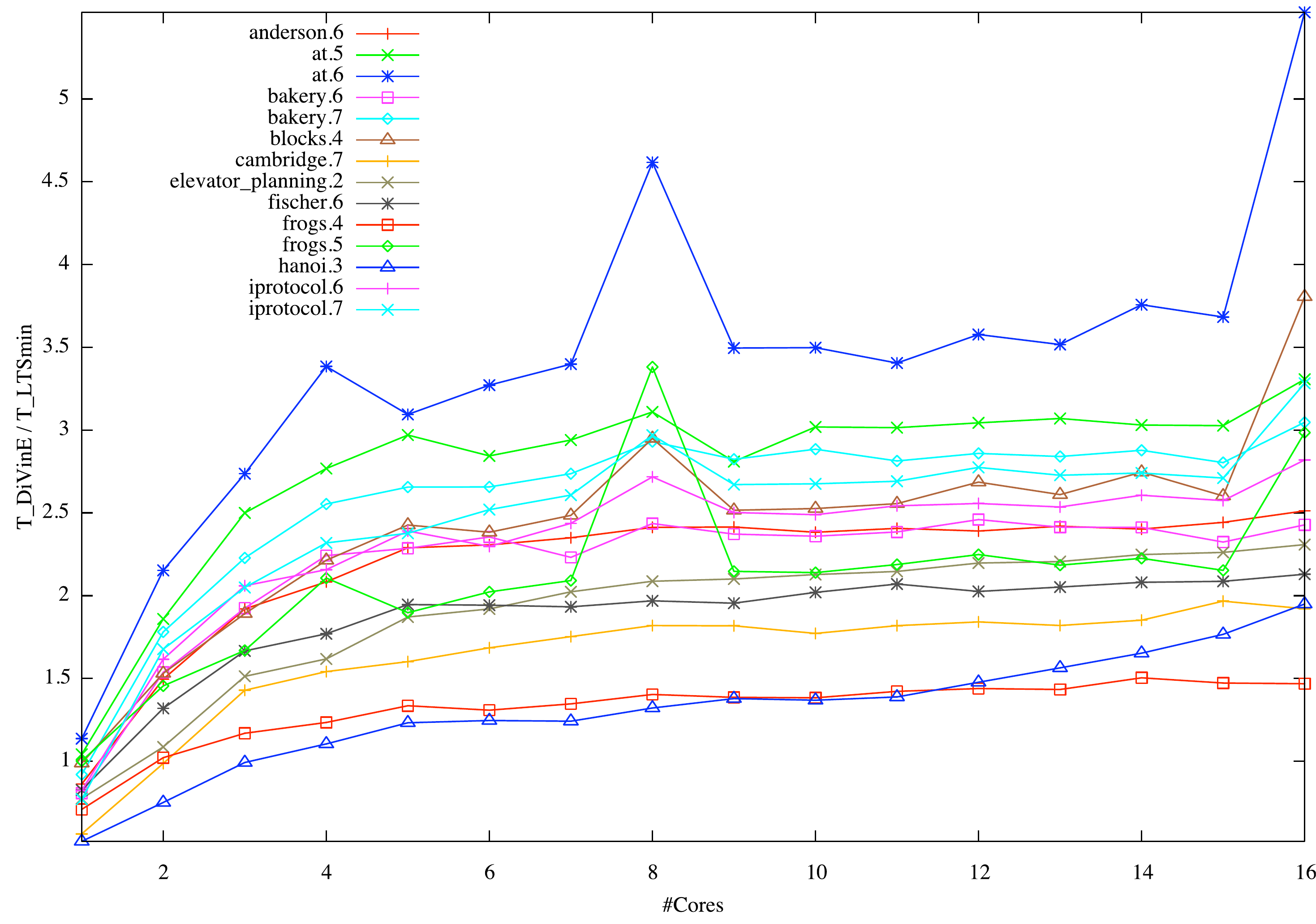} 
 
  \includegraphics[width=\textwidth]{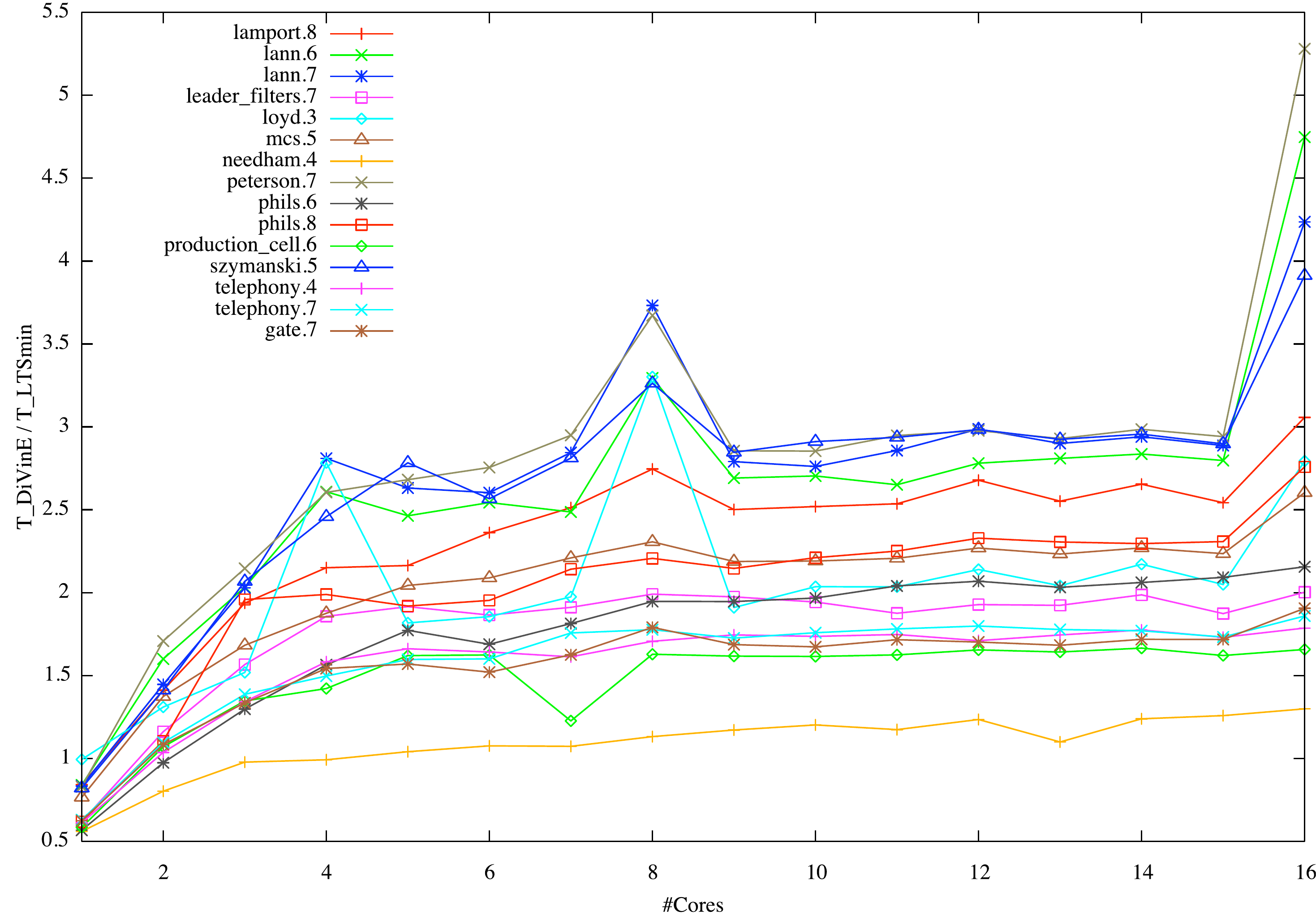} 
  \caption{Runtimes of BEEM models with LTSmin and DiVinE 2}
\label{fig:reldiv}
\end{center}
\end{figure}

\begin{figure}[h!]
\begin{center} 
 \includegraphics[width=\textwidth]{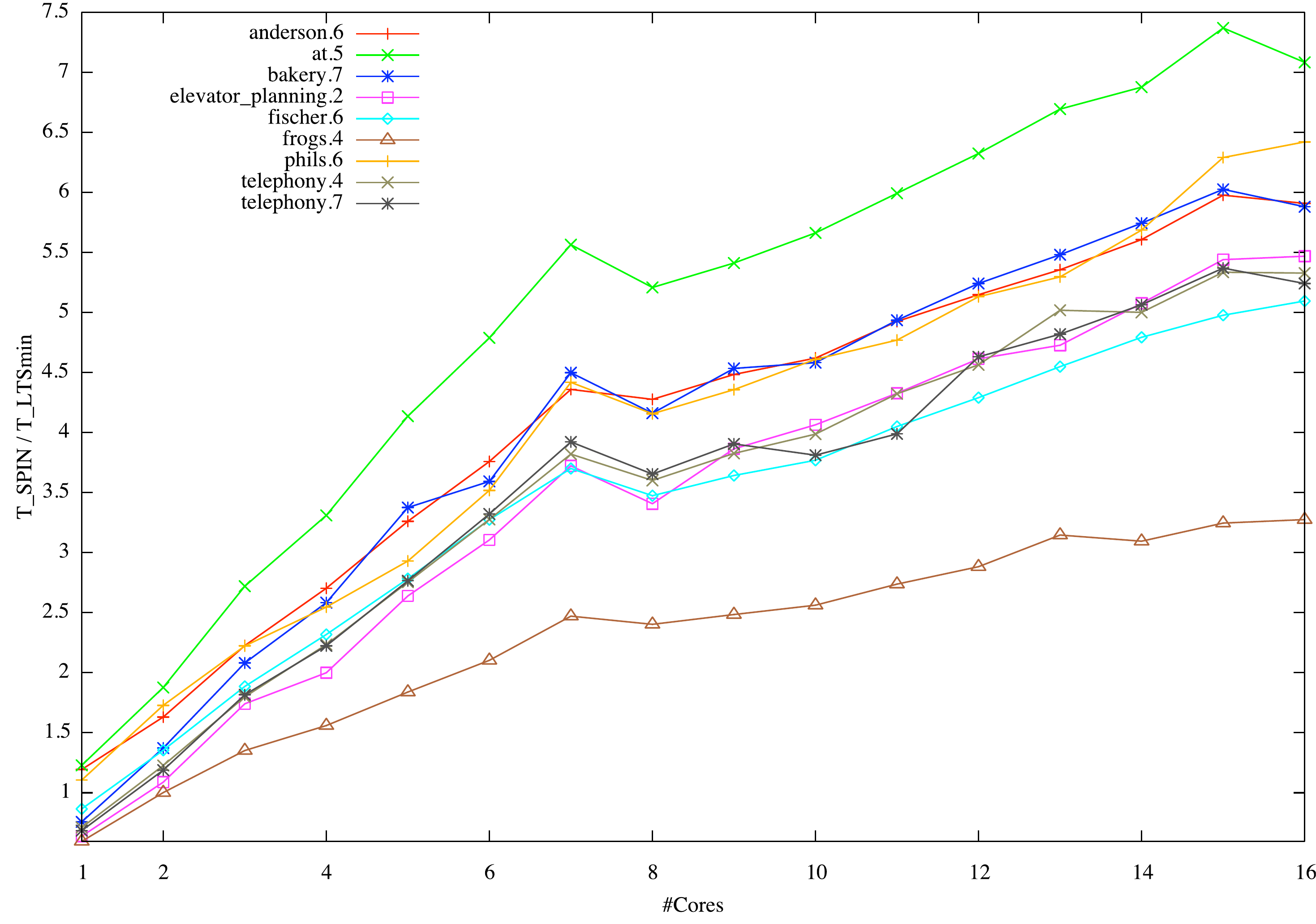} 
  \caption{Runtimes of BEEM models with LTSmin and SPIN}
\label{fig:relspin}
  \end{center}
\end{figure}

LTSmin is slower in the sequential cases. We verified that this behavior is caused by allocation-less 
hash table; with smaller hash table sizes the sequential runtimes match those of DiVinE. However, we 
did not bother optimizing these results, because with two cores LTSmin is already at least as fast as 
DiVinE.

To show all models in a similar plot is hardly feasible. Therefore, Fig.~\ref{fig:reldiv} and 
\ref{fig:relspin} compare the relative runtimes per model of two model checkers: $
\frac{T_{\mathrm{DiVinE}}}{T_{\mathrm{LTSmin}}}$ and $\frac{T_{\mathrm{SPIN}}}
{T_{\mathrm{LTSmin}}}$. Fig.~\ref{div2} in the appendix shows the speedups measured with LTSmin 
and DiVinE. We attribute the difference in scalability to the extra synchronization points needed for 
the interprocess communication by DiVinE. Remember that the model checker uses static state 
space partitioning, meaning that most successor states are queued at other cores. Another 
disadvantage of DiVinE, is that it uses a management thread. This causes the regression at 
16~cores.

SPIN shows inferior scalability even though it also uses a shared hash table and load balancing based 
on stack slicing. We can only guess that the spinlocks SPIN uses in its hash table (region locking) 
are not as efficient as a lockless hash table. However, we had far better results even with the slower 
\texttt{pthread} locks. It might also be that the stack slicing algorithm does not have a consistent 
granularity, because it uses the irregular search depth as a time unit (using the terms from Sec.~
\ref{sec:preliminaries}: $T(work(P,depth))$).

\begin{figure}[h!]
\begin{center} 
   \subfloat[][]{
\includegraphics[width=0.5\textwidth]{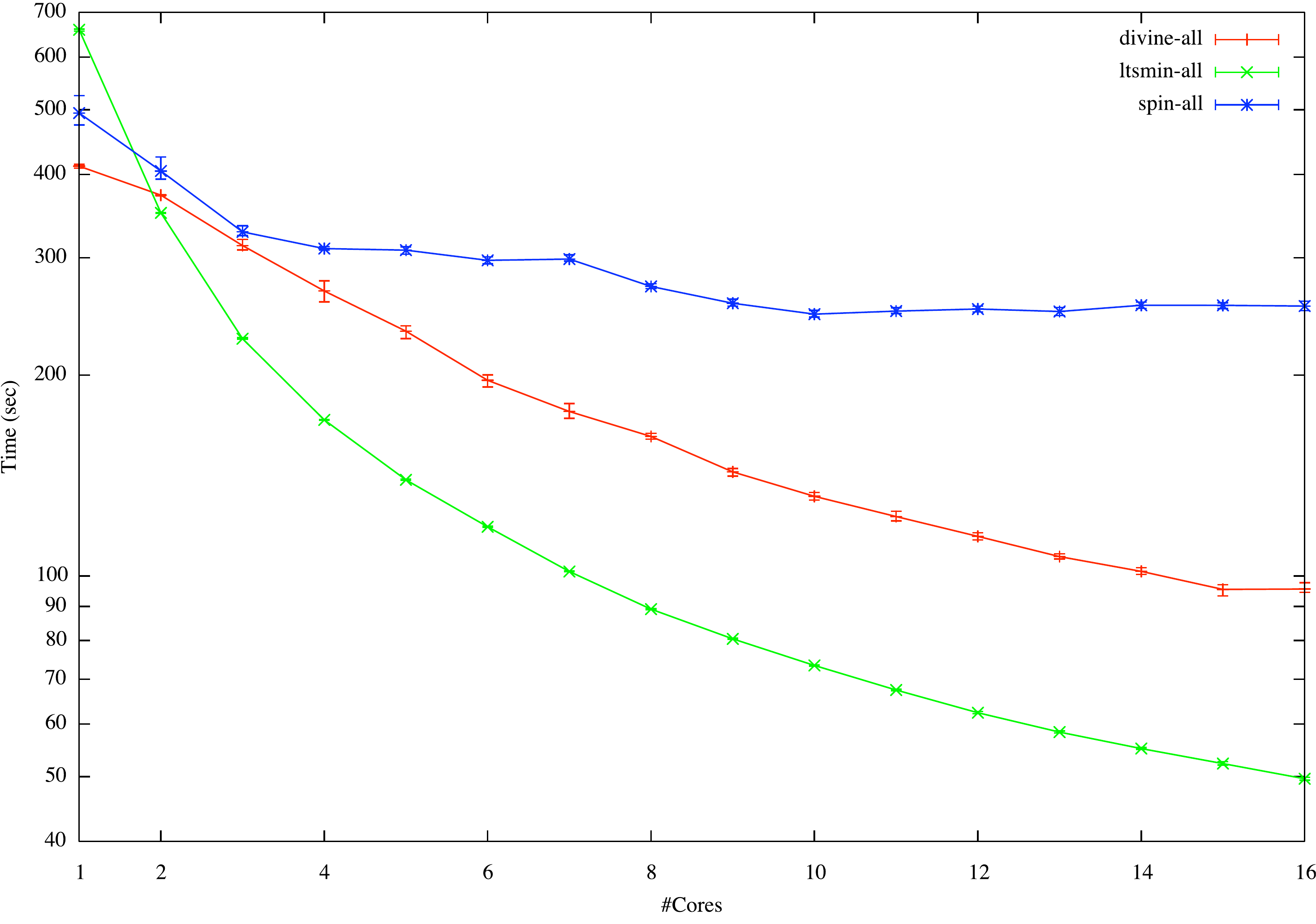}}
 \subfloat[][]{
\includegraphics[width=0.5\textwidth]{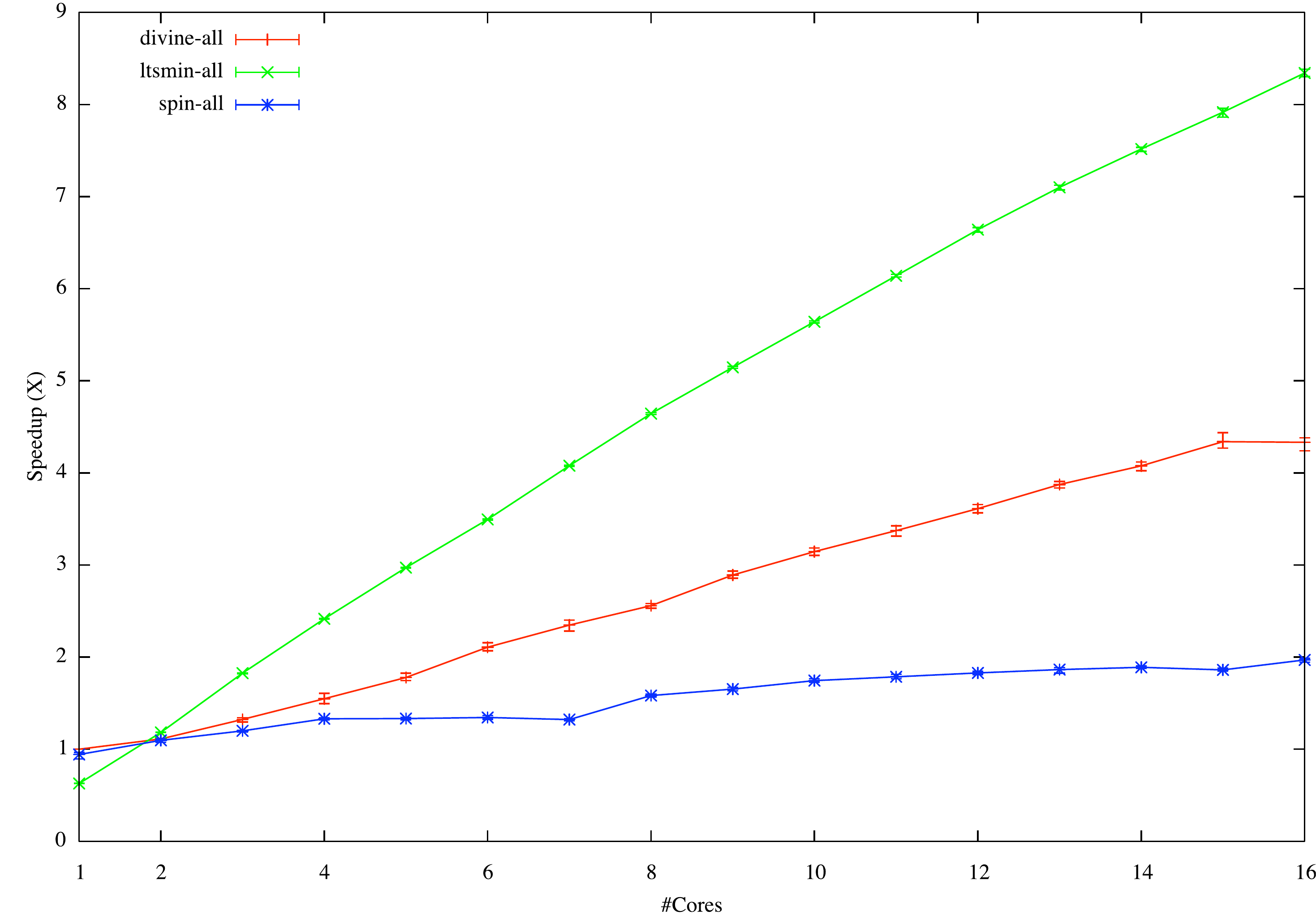}}
\caption{Total runtime and speedups of Spin, Divine 2 and LTSmin-mc}
\label{dvevsdiv2}
 \end{center}
\end{figure}

Fig.~\ref{dvevsdiv2} shows the average times (a) and speedups (b) over all models and for all model 
checkers. These are only the marked models in Table~\ref{tab:models}, because they have the same 
state count in all tested model checkers.

\subsection{Shared Storage Parameters}

\begin{figure}[h!]
\begin{center}
\includegraphics[width=.7\textwidth]{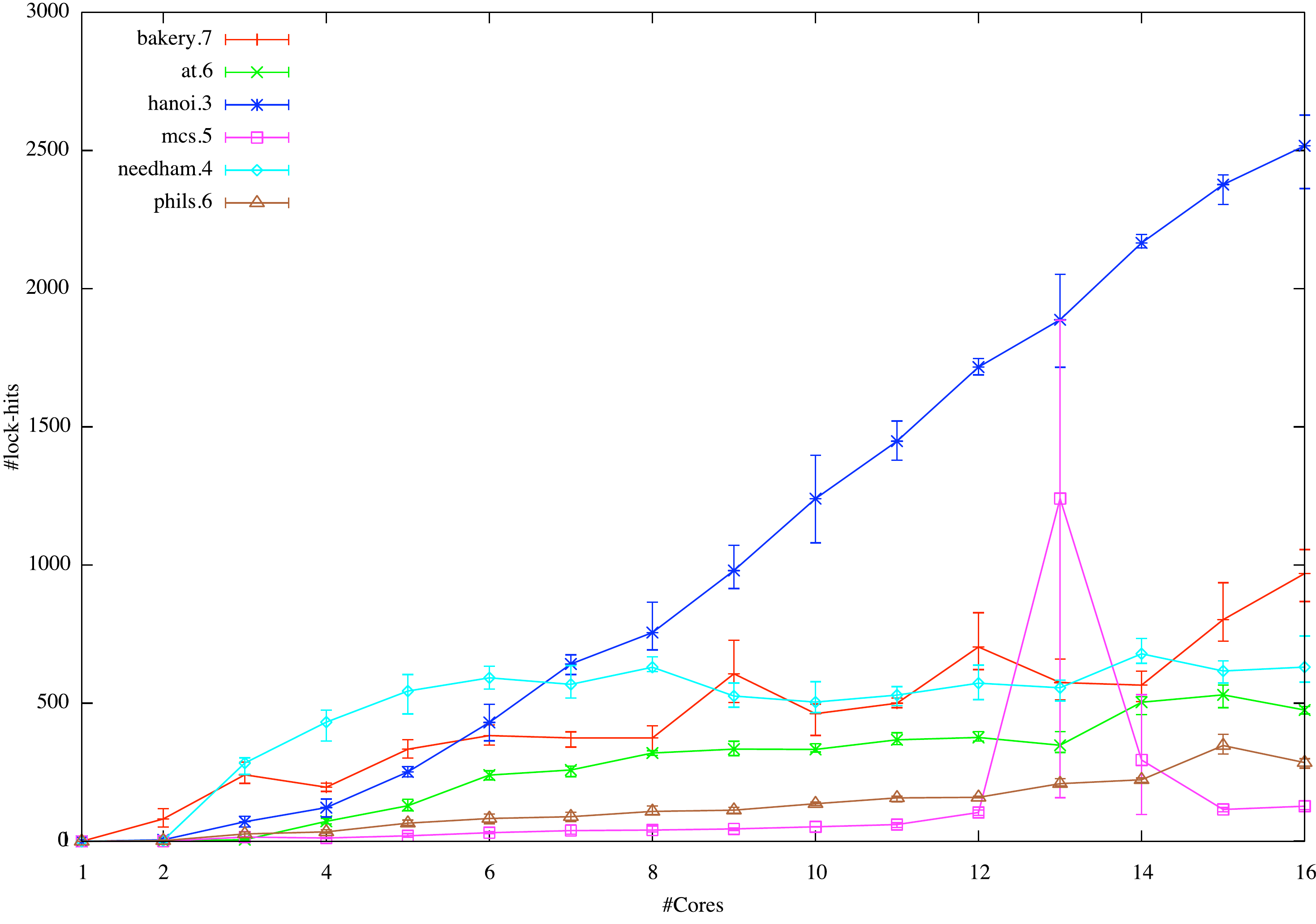}
\caption{Times the algorithm ``locks''}
\label{fig:waits}
\end{center}
\end{figure}

To verify our claims about the hash table design, we collected internal measurements and synthetic 
benchmarks. First, we looked at the number of times that the write-busy ``lock'' was hit. Fig.~
\ref{fig:waits} plots the lock hits against the number of cores for several different sized models. For 
readability, only the highest and most interesting models where chosen. Even for the largest models 
(at.6) the number of locks is a very small fraction of the number of \textit{find-or-put} calls: 160M 
(equal to the number of transitions).  

\begin{figure}[h!]
\begin{center}
\subfloat[][]{
  \includegraphics[width=.8\textwidth]{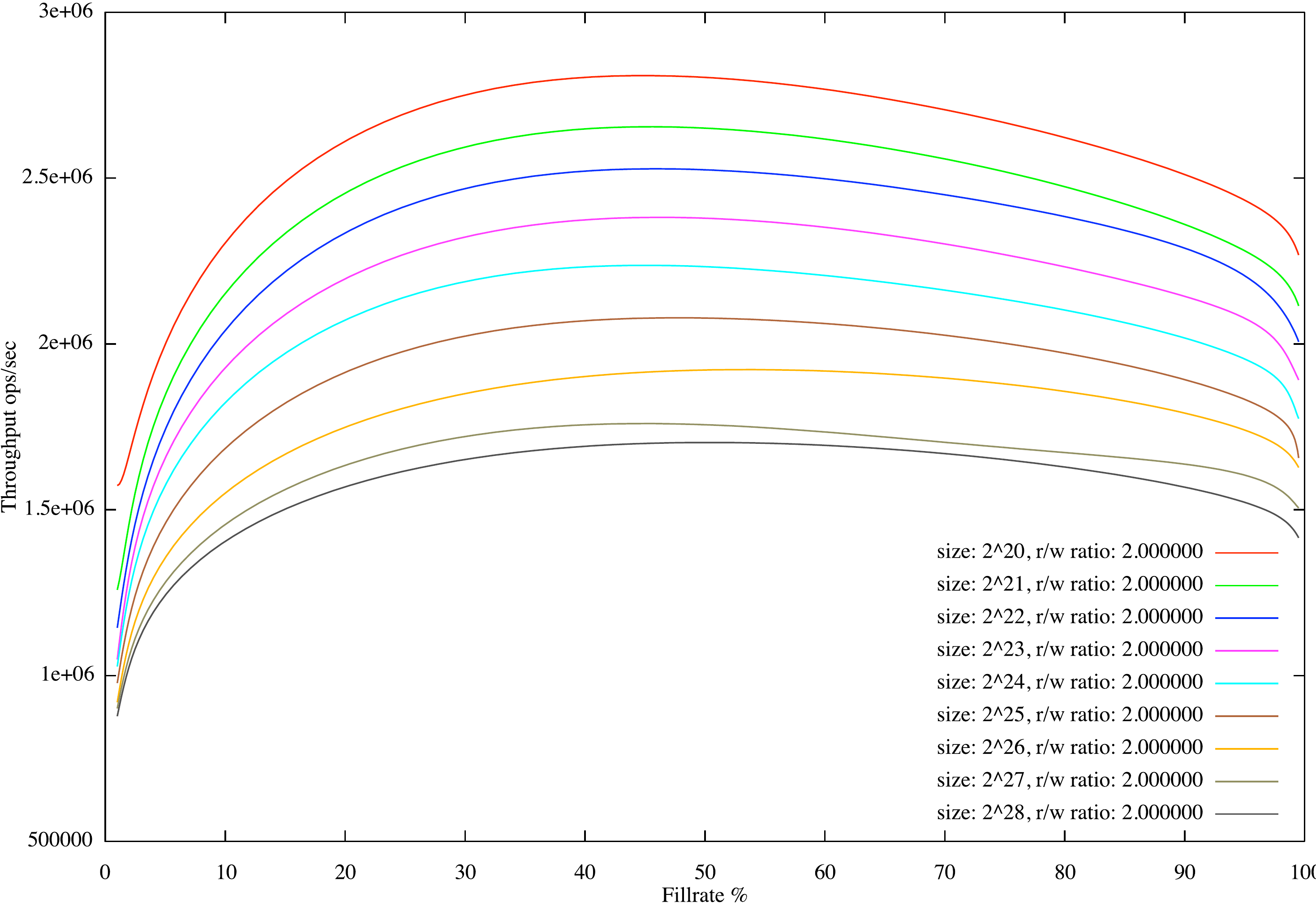} \label{fig:size}}

\subfloat[][]{
  \includegraphics[width=.8\textwidth]{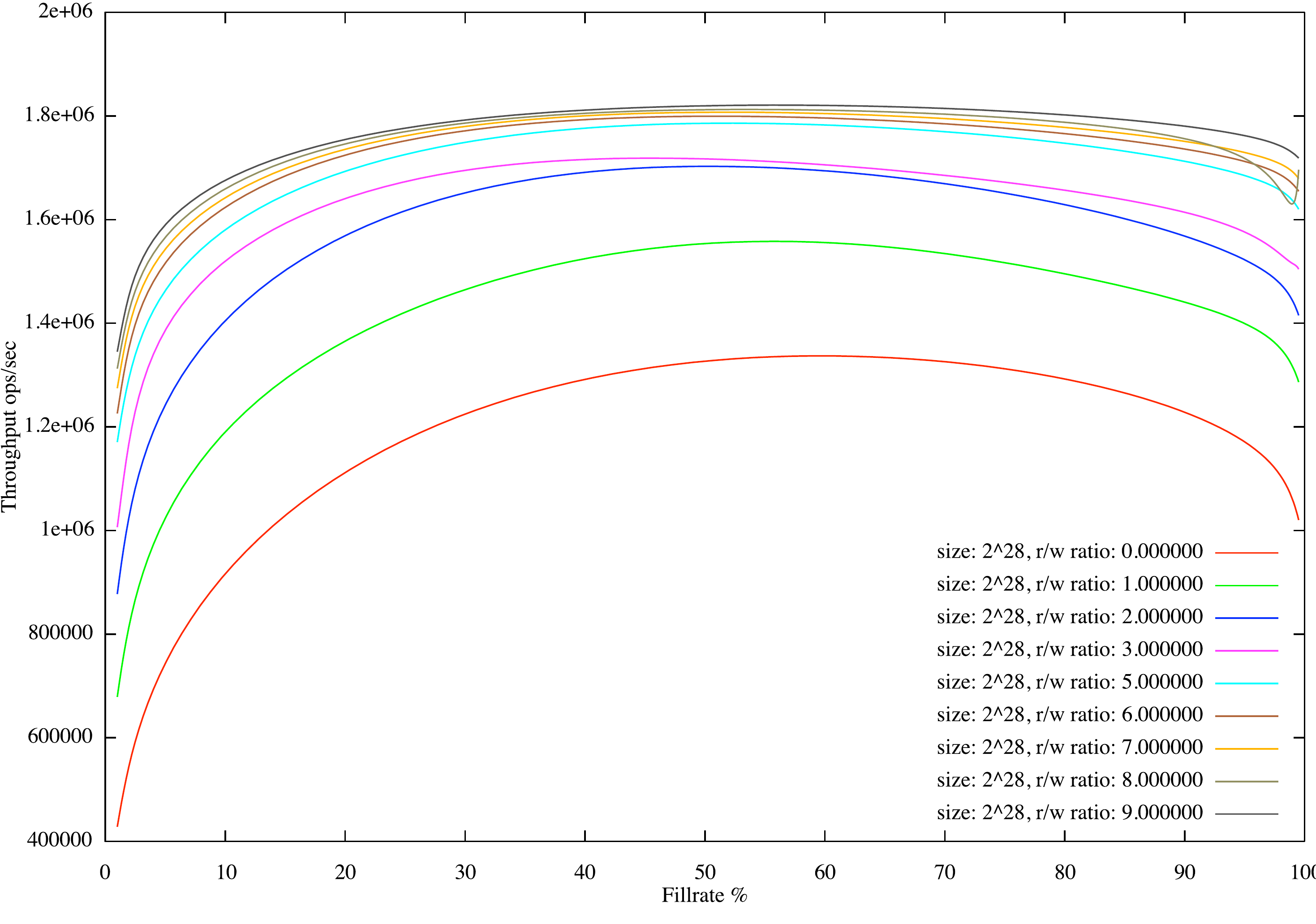} \label{fig:rw}}
\end{center}
\caption{Effect of fill-rate and size/rw-write on average throughput}
\label{fig:complexities}
\end{figure}

 We measured how the average throughput of Alg.~\ref{alg:findorput} (number of \textit{find-or-put} 
calls) is effected by the fill-rate, the table size and the read write ratio. Fig.~\ref{fig:complexities} 
shows measurements done with synthetic input data that simulates one ratio and n (random) reads on 
the hash table. Both figures show that average throughput remains largely unaffected by high fill-rate, 
even up to 99.9\%. This shows that the asymptotic time complexity of open-addressing hash tables 
poses little real problems in practice. An observable side effect of large hash tables, is lower 
throughputs for low fill rates due to more cache misses. Our hash table amplifies this because it uses 
a pre-allocated data array and no pointers. This explains the sequential performance difference 
between DiVinE, SPIN and our model checker. The following section will follow up on this.

We also measured the effect of varying the vector size (not in the figures) and did not find any 
noticeable change in the figures (except for the expected lower throughputs). This showed us that 
hash memoization and a separate data array do a good job. 
At this point, many other questions can be asked and answered by further investigation.  These would 
be out of scope now, but are on our list of future work.



%% file: 5_conclusion.tex
\section{Discussion and Conclusions}

We designed a hash table suitable for application in reachability analysis. We implemented it as part of 
a model checker together with different exploration algorithms (pseudo BFS and pseudo DFS) and 
explicit load-balancing. We demonstrated the efficiency of the complete solution, by comparing the 
absolute speedups compared to the SPIN 5.2.4 and the DiVinE 2.2 model checkers, leading tools in 
this field. We claim two times better scalability than DiVinE and five times better than SPIN \emph{on 
average}. We also investigated the properties of the hash table with regard to its behavior under 
different fill~rates and found it to live up to the requirements we imposed.

\paragraph*{Limitations.}
Without the use of pointers the current design cannot cope with variably sized state vectors. In our 
model checker, this does not pose a problem because states are always represented by a vector of a 
static length. Our model checker LTSmin\footnote[\value{foot}]{\url{http://fmt.cs.utwente.nl/tools/
ltsmin/}} can handle different front-ends. It connects to DiVinE-cluster, DiVinE 2.2, PROMELA (via the 
\textsc{NipsVM}~\cite{so62050}), mCRL, mCRL2 and ETF (internal symbolic representation of state 
spaces). Some of these input languages require variably sized vectors (NIPS). This is solved by an 
initial exploration, that continues until a stable vector is found. So far, this limitation did not pose a 
problem.

\addtocounter{foot}{1}

The results in the sequential case proved around 20\% slower than DiVinE 2.2. We can explain this 
because we have one extra level of indirection (function calls) to abstract from the language front-end. 
But, this is not the only reason. It turns out that for large models we are actually as fast as DiVinE. 
Small models, however, underperform compared to DiVinE (up to 50\% as can be seen in the figures 
with absolute speedups). The difference here is caused by the pointer-less and allocation-less data 
structure, which simply introduces more cache misses with low fill rates. When we embrace pointers 
and allocation this could be remedied, but the question is whether such a solution will still scale, 
because cache hits can cause cache line sharing and thus extra communication in parallel operation.

\paragraph*{Discussion.}
We make several observations based on the results presented here:
\begin{itemize}

\item Centralized state storage scales better and is more flexible. It allows for pseudo DFS (like the 
stack slicing algorithm), but can also facilitate explicit load balancing algorithms. The latter opens the 
potential to exploit heterogenous systems. Early results with load balancing showed a few percent 
overhead compared to static load balancing.

\item Performance-critical parallel software needs to be adapted to modern architectures (steep 
memory hierarchies). An indication of this can be seen in the performance difference between DiVinE, 
SPIN and LTSmin. DiVinE uses an architecture which is directly derived from distributed model 
checking and the goal of SPIN was for ``these new algorithms [\textellipsis] to interfere as little as 
possible with the existing algorithms for the veriÞcation of safety and liveness properties''~
\cite{Holzmann20083}. With LTSmin, on the other hand, we had the opportunity to tune our design to 
the architecture of the machine. We noticed that avoiding cache line sharing and keeping a simple 
design was instrumental to handle the system's complexities. 

\item Holzmann made the conjecture that optimized sequential code does not scale well~\cite{HB07}.  
In contrast, our parallel implementation is faster in absolute numbers and also exhibits excellent 
scalability.
We suspect that the (entirely commendable) design choice of SPIN's multi-core implementation to 
support most of SPIN's existing features is detrimental to scalability.  In our experience, parallel 
solutions work best if they are tailored to each individual problem.

\item Scalable enumerative reachability is a good starting point for future work on multi-core (weak) LTL 
model checking, symbolic exploration and space-efficient explicit exploration.
\end{itemize}

\paragraph*{Future work.}
As mentioned above, this work has been conducted in the broader context of the LTSmin model 
checker. The goal of LTSmin is to deliver language independent model checking algorithms without a 
performance penalty. We do this by finding a suitable representation for information from the language 
engine that is normally used for optimization. By means of a \emph{dependency matrix} \cite{TR-
CTIT-09-30} this information is preserved and utilized for symbolic exploration, (distributed) reachability 
and (distributed) state space minimization with bisimulation \cite{BO05}. In this work, we showed 
indeed that we do not have to pay a performance penalty for language independent model checking.

for multi-core mCRL algorithms by using POSIX shared memory to accommodate the inherently 
sequential implementation of mCRL.

By exploring the possible solutions and gradually improving this work, we found a wealth of variables 
hiding in the algorithms and the models of the BEEM database. As can be seen from the figures, 
different models show different scalability. By now we have some ideas where these differences come 
from. For example, an initial version of the exploration algorithm employed static load balancing by 
means of an initial BFS exploration and handing of the states from the last level to all threads. Several 
models where insensitive to this static approach, others, like hanoi.x and frogs.x, are very sensitive 
due to the form of there state space. Dynamic load balancing did not come with a noticeable 
performance penalty for the other models, but hanoi and frogs are still in the bottom of the figures. 
There are still many options open to improve shared memory load balancing and remedy this. 

We also experimented with space-efficient storage in the form of a tree compression. Results, also 
partly based on algorithms presented here, are very promising and we intend to follow up on that.


%% file: 0_LLDBS.bbl
\begin{thebibliography}{10}

\bibitem{balanced}
Yossi Azar, Andrei~Z. Broder, Anna~R. Karlin, and Eli Upfal.
\newblock Balanced allocations.
\newblock {\em SIAM J. Comput.}, 29(1):180--200, 2000.

\bibitem{BBR07}
Ji\v{r}\'{\i} Barnat, Lubo\v{s} Brim, and P.~Ro\v{c}kai.
\newblock Scalable multi-core {LTL} model-checking.
\newblock In {\em Model Checking Software}, volume 4595 of {\em LNCS}, pages
  187--203. Springer, 2007.

\bibitem{BBR09}
Ji\v{r}\'{\i} Barnat, Lubo\v{s} Brim, and Petr Ro\v{c}kai.
\newblock {DiVinE 2.0: High-Performance Model Checking}.
\newblock In {\em 2009 International Workshop on High Performance Computational
  Systems Biology (HiBi 2009)}, pages 31--32. IEEE Computer Society Press,
  2009.

\bibitem{Barnat200879}
Ji\v{r}\'{\i} Barnat and Petr Ro\v{c}kai.
\newblock Shared hash tables in parallel model checking.
\newblock {\em Electronic Notes in Theoretical Computer Science}, 198(1):79 --
  91, 2008.
\newblock Proceedings of the 6th International Workshop on Parallel and
  Distributed Methods in verifiCation (PDMC 2007).

\bibitem{TR-CTIT-09-30}
S.~C.~C. Blom, J.~C. van~de Pol, and M.~Weber.
\newblock Bridging the gap between enumerative and symbolic model checkers.
\newblock Technical Report TR-CTIT-09-30, University of Twente, Enschede, June
  2009.

\bibitem{BO05}
Stefan Blom and Simona Orzan.
\newblock A distributed algorithm for strong bisimulation reduction of state
  spaces.
\newblock {\em STTT}, 7(1):74--86, 2005.

\bibitem{Brim:PDMC:2006}
Lubo\v{s} Brim.
\newblock Distributed verification: Exploring the power of raw computing power.
\newblock In Lubo\v{s} Brim, Boudewijn Haverkort, Martin Leucker, and Jaco
  van~de Pol, editors, {\em Formal Methods: Applications and Technology},
  volume 4346 of {\em Lecture Notes in Computer Science}, pages 23--34.
  Springer, August 2006.

\bibitem{CC07}
Cliff Click.
\newblock Performance myths exposed.
\newblock Talk at JavaOne 2007, 2007.

\bibitem{terminator}
Byron Cook, Andreas Podelski, and Andrey Rybalchenko.
\newblock Terminator: Beyond safety (tool paper).

\bibitem{Cormen}
Thomas~H. Cormen, Charles~E. Leiserson, Ronald~L. Rivest, and Clifford Stein.
\newblock {\em Introduction to Algorithms, Third Edition}.
\newblock The MIT Press, 3 edition, September 2009.

\bibitem{JavaSpec}
James Gosling, Bill Joy, Guy Steele, and Gilad Bracha.
\newblock {\em The Java Language Specification, Third Edition}.
\newblock Addison-Wesley Longman, Amsterdam, 3 edition, June 2005.

\bibitem{speedup}
D.~P. Helmbold and C.~E. McDowell.
\newblock Modeling speedup (n) greater than n.
\newblock {\em IEEE Trans. Parallel Distrib. Syst.}, 1(2):250--256, 1990.

\bibitem{amp}
Maurice Herlihy and Nir Shavit.
\newblock {\em The Art of Multiprocessor Programming}.
\newblock Morgan Kaufmann, March 2008.

\bibitem{HNM09}
Maurice Herlihy, Nir Shavit, and Moran Tzafrir.
\newblock Hopscotch hashing.
\newblock {\em Distributed Computing}, pages 350--364, 2008///.

\bibitem{Holzmann20083}
Gerard~J. Holzmann.
\newblock A stack-slicing algorithm for multi-core model checking.
\newblock {\em Electronic Notes in Theoretical Computer Science}, 198(1):3 --
  16, 2008.
\newblock Proceedings of the 6th International Workshop on Parallel and
  Distributed Methods in verifiCation (PDMC 2007).

\bibitem{HB07}
Gerard~J. Holzmann and Dragan Bo\v{s}nacki.
\newblock The design of a multicore extension of the spin model checker.
\newblock {\em IEEE Trans. Softw. Eng.}, 33(10):659--674, 2007.

\bibitem{JE10}
Shahid Jabbar and Stefan Edelkamp.
\newblock Parallel external directed model checking with linear i/o.
\newblock {\em Verification, Model Checking, and Abstract Interpretation},
  pages 237--251, 2006///.

\bibitem{litwin80}
Witold Litwin.
\newblock Linear hashing: a new tool for file and table addressing.
\newblock In {\em VLDB '1980: Proceedings of the sixth international conference
  on Very Large Data Bases}, pages 212--223. VLDB Endowment, 1980.

\bibitem{MP09}
Michael Monagan and Roman Pearce.
\newblock Parallel sparse polynomial multiplication using heaps.
\newblock In {\em ISSAC '09: Proceedings of the 2009 international symposium on
  Symbolic and algebraic computation}, pages 263--270, New York, NY, USA, 2009.
  ACM.

\bibitem{PFR01}
Rasmus Pagh and Flemming~Friche Rodler.
\newblock Cuckoo hashing.
\newblock {\em Journal of Algorithms}, 51(2):122 -- 144, 2004.

\bibitem{beem}
R.~Pel\'anek.
\newblock Beem: Benchmarks for explicit model checkers.
\newblock In {\em Proc. of SPIN Workshop}, volume 4595 of {\em LNCS}, pages
  263--267. Springer, 2007.

\bibitem{opaddrht}
Chris Purcell and Tim Harris.
\newblock Non-blocking hashtables with open addressing.
\newblock {\em Distributed Computing}, pages 108--121, 2005///.

\bibitem{RKV10}
V.~Rao and Vipin Kumar.
\newblock Superlinear speedup in parallel state-space search.
\newblock {\em Foundations of Software Technology and Theoretical Computer
  Science}, pages 161--174, 1988///.

\bibitem{RK88}
V.~Rao and Vipin Kumar.
\newblock Superlinear speedup in parallel state-space search.
\newblock {\em Foundations of Software Technology and Theoretical Computer
  Science}, pages 161--174, 1988///.

\bibitem{sandersthesis}
P~Sanders.
\newblock Lastverteilungsalgorithmen fur parallele tiefensuche. number 463.
\newblock In {\em in Fortschrittsberichte, Reihe 10. VDI}. Verlag, 1997.

\bibitem{delight}
Henry~S. Warren.
\newblock {\em Hacker's Delight}.
\newblock Addison-Wesley Longman Publishing Co., Inc., Boston, MA, USA, 2002.

\bibitem{so62050}
M.~Weber.
\newblock An embeddable virtual machine for state space generation.
\newblock In D.~Bosnacki and S.~Edelkamp, editors, {\em Proceedings of the 14th
  International SPIN Workshop}, volume 2595 of {\em Lecture Notes in Computer
  Science}, pages 168--186, Berlin, 2007. Springer Verlag.

\end{thebibliography}
